\documentclass[12pt]{article}
\usepackage{amsmath}
\usepackage{amssymb}
\usepackage{graphicx}
\usepackage{ifthen}
\usepackage{verbatim}
\usepackage{psfrag}

\numberwithin{equation}{section}

\begin{document}

\begin{flushright}
\texttt{hep-th/0604141}\\
\texttt{OU-HET 559}\\
April 2006
\end{flushright}
\bigskip
\bigskip
\begin{center}
{\Large \textbf{Supersymmetric Gauge Theories with Matters,}}\\
{\Large \textbf{ Toric Geometries and Random Partitions}}\\

\end{center}

\vspace{20mm}
\begin{center}
Yui Noma\footnote{E-mail: yuhii@het.phys.sci.osaka-u.ac.jp}

{\small 
\textit{Department of Physics, Graduate School of Science, 
Osaka University,\\
Toyonaka, Osaka 560-0043, Japan\\}}
\end{center}

\vspace{40mm}
\begin{abstract}
We derive
 the relation
 between 
 the Hilbert space 
 of certain geometries
 under the Bohr-Sommerfeld quantization
 and 
 the perturbative prepotentials
 for the supersymmetric
 five-dimensional $SU(N)$ gauge theories
 with massive fundamental matters
 and with one massive adjoint matter.

The gauge theory with one adjoint matter
 shows interesting features.
A five-dimensional generalization 
 of Nekrasov's partition function
 can be written as 
 a correlation function
 of two-dimensional chiral bosons
 and as a partition function of 
 a statistical model of partitions.
From a ground state of the statistical model
 we reproduce
 the polyhedron which characterizes
 the Hilbert space.
\end{abstract}
\newpage

\section{Introduction}
Gauge/Gravity correspondences 
 are fascinating properties
 suggested from the string theory.
Recently, interesting correspondences
 were found \cite{MNTT1}
 \cite{MNTT2}
 \cite{MNNT} \cite{amoeba}
 which relate
 the five-dimensional $SU(N)$ gauge theory
 compactified on a circle
 with eight super-charges,
 a certain toric variety
 called a local $SU(N)$ geometry 
 and
 a certain statistical model of partitions
 which is called a random plane partition model.
In this paper we generalize 
 the correspondence to the case of 
 the gauge theory
 with matters.

The four-dimensional gauge theory
 with eight super-charges 
 was solved by Seiberg and Witten
 \cite{seiberg-witten}
 and 
 its low energy effective action
 is determined by a holomorphic function
 which is called a prepotential.
The prepotential consists of
 perturbative terms and non-perturbative terms.
Nekrasov and Okounkov solved the theory without matters
 by using a localization technique and a random partition model
 \cite{nek okoun}.
They generalized
 the partition function
 to the case of the gauge theory with fundamental matters,
 with one adjoint matter 
 and the five-dimensional gauge theory on a circle.
The partition functions
 are called 
 Nekrasov's partition functions.
Each of them is factorized to
 two parts,
 which are called 
 the perturbative part and the instanton part.
The perturbative (non-perturbative)
 prepotential
 is obtained from
 the perturbative (instanton) part of
 the partition function.

Seiberg et al.
 determined 
 the perturbative 
 prepotentials for the five-dimensional gauge theories
 with or without matters \cite{5d seiberg}.
The perturbative prepotential for the $SU(N)$ gauge theory
 is as follows:
\begin{eqnarray}
\mathcal{F}^{pert}(a_r; m_f)
=\frac{1}{2g_{YM}^2} \sum_{r=1}^{N} a_r^2
 +\frac{k_{\mbox{\tiny C.S.}}}{6}
 \sum_{r=1}^{N} a_r^3
 + \frac{1}{12}\left(
 2 \sum_{r>s} |a_r-a_s|^3
 -\sum_{f}\sum_{{\bf w} \in {\bf W}_f} |{\bf w}\cdot a + m_f|^3
\right),
\end{eqnarray}
where
 $a_r$ are vev. of the adjoint scalars,
 $g_{YM}$ is the gauge coupling constant,
 $f$ labels the matters,
 $m_f$ is the mass of the $f$-th matter,
 ${\bf W}_f$ are the weights 
 for the $f$-th matter,
 and
 $k_{\mbox{\tiny C.S.}}$ is the Chern-Simons
 coupling constant.
For each theory,
 the cubic terms in the perturbative prepotential
 are derived from 
 triple-intersection numbers of divisors
 of a certain Calabi-Yau manifold.
However triple-intersection numbers of 
 non-compact divisors are subtle.
The author derived
 the cubic and quadratic terms
 in
 the perturbative prepotential
 for the theory without matters
 by calculating volume of a certain polyhedron
 \cite{MNNT}.

If the geometry is a toric variety,
 it is described 
 by a polyhedron $\mathcal{P}$
 defined on a lattice $M^{\vee}$
 \cite{toric}.
Integer lattice points
 in $\mathcal{P}\cap M^{\vee}$ can be seen as
 physical states
 of the Hilbert space for
 the geometry
 under the Bohr-Sommerfeld quantization.
The cardinality of $\mathcal{P}\cap M^{\vee}$
 becomes the dimension of
 the Hilbert space.
For the local $SU(N)$ geometry,
 which is non-compact and 
 can be seen as 
 an ALE space fibred over $\mathbb{CP}^1$,
 the cardinality
 is infinite 
 and
 the dimension is naively infinite.
Nevertheless it is
 regularized 
 by using a complement of $\mathcal{P}$,
 which is a deviation
 from a singular limit.
With a suitable identification
 between geometric parameters and
 gauge theoretic parameters,
 the perturbative prepotential 
 for the five-dimensional gauge theory without matters
 emerges
 from the dimension.

$\mathcal{P}$ can be introduced by 
 means of a Weil divisor $D$.
When the toric variety is compact 
 and 
 the polyhedron 
 is enough large
 for the lattice to be
 regarded as continuous,
 the self-intersection numbers
 of $D$ and 
 the volume of $\mathcal{P}$
 are related as follows:
\begin{eqnarray}
\frac{1}{d !} D^d = \mbox{Vol} (\mathcal{P}),
 \label{eq;vol and inters}
\end{eqnarray}
 where $d$ is the dimension of the variety
 and $D^d$ 
 denotes $d$-times self-intersection
 of $D$.
However for non-compact toric varieties 
 there are no equation like (\ref{eq;vol and inters}).

The instanton part of Nekrasov's partition functions
 is obtained from
 the amplitude of the topological A-model strings
 on the local geometry
 by using topological vertexes
 \cite{pure inst top ver}
 \cite{iqbal 5d}.
The topological vertex can be seen as
 a random plane partition model \cite{rpp1}.
The perturbative and instanton part of
 Nekrasov's partition function
 for the five-dimensional gauge theory
 without matters
 are derived
 from the random plane partition model \cite{MNTT1}.
The polyhedron for 
 the local geometry
 is reproduced
 from a ground state
 of the random plane partition model  \cite{MNNT}.

In this paper,
 we generalize
 the triality,
 between
 gauge theories, geometries and statistical models.
We reconstruct 
 the perturbative prepotentials
 for the five-dimensional $SU(N)$ gauge theories
 with fundamental matters and
 with one adjoint matter
 from the dimension
 of the Hilbert spaces of certain geometries.
We express 
 a five-dimensional generalization
 of Nekrasov's partition function
 for the $U(1)$
 gauge theory
 with one adjoint matter
 as
 a correlation function of two-dimensional 
 free chiral bosons
 and
 as 
 a partition function of 
 a certain statistical model of partitions.
From this statistical model
 we can reproduce the polyhedron
 for the 
 $SU(N)$ gauge theory
 with one adjoint matter.

This paper is organized as follows.
In Section \ref{sec;dim count fun},
 we derive the relation between
 certain geometries
 and perturbative prepotentials for 
 the five-dimensional $SU(N)$ gauge theories
 with matters in the fundamental representation.
In Section \ref{sec;dim count adj},
 we derive the relation between a certain geometry
 and the five-dimensional 
 gauge theory with one adjoint matter.
In Section \ref{sec;cft and N=1*},
 we study several properties
 of Nekrasov's partition function for
 the five-dimensional gauge theory
 with one adjoint matter.
In particular,
 we show that
 it is expressed
 as 
 a correlation function 
 of two-dimensional free chiral bosons.
In Section \ref{sec;rpp},
 we rewrite the correlation function
 as
 a partition function of
 a statistical model of partitions
 and 
 reproduce the polyhedron for
 the geometry
 from a ground state of the statistical model.
Relations between
 fermions and partitions
 are in appendix  \ref{sec;fermion}.

\section{
Perturbative prepotentials with fundamental matters
 from polyhedrons}
\label{sec;dim count fun}
\subsection{The case of $N_F=0$}
\label{sec;dim count no matter}

Our goal in this section
 is to describe
 the relation 
 between the Hilbert space
 for a certain toric variety
 and the perturbative prepotential for
 the five-dimensional $SU(N)$ gauge theory
 with
 massive matters in the fundamental representation.
We conveniently start
 with the case of
 no matters.
In this case,
 the relevant geometry is
 called
 the local $SU(N)$ geometry
 as described in \cite{MNNT}.
It is
 a non-compact Calabi-Yau manifold, 
 a three-dimensional toric variety
 and seen as an ALE space fibred over $\mathbb{CP}^1$.
The fibration is characterized by an integer 
 $k_{\mbox{\tiny C.S.}} \in [1, N]$,
 which is called framing.
The geometry is quantized
 by the Bohr-Sommerfeld quantization.
States in the Hilbert space
 turn to be represented by
 integer lattice points in
 a certain polyhedron.
The dimension of the Hilbert space 
 approximates to
 the volume
 of this polyhedron
 as the string coupling constant
 $g_{st}$ goes to $0$.
By a certain identification 
 between geometric parameters 
 and gauge theoretic parameters,
 the volume yields
 the perturbative prepotential for the 
 five-dimensional gauge theory with no matters.

Recall that
 in quantum mechanics
 the Bohr-Sommerfeld quantization 
 is carried out by imposing
 a condition 
 on classical paths in a phase space.
In the case of  
 a particle moving in $d$-dimensional space,
 the condition
 for the Bohr-Sommerfeld orbits $C$
 is 
\begin{eqnarray}
\frac{1}{h}\oint_{C}
\sum_{i=1}^{d} 
 p_i \, dq^i \in \mathbb{Z},
\end{eqnarray}
where
 $q^i$ is
 the $i$-th component of
 the coordinate of the particle,
 $p_i$ is the $i$-th component of
 its momentum
 and $h$ is the Planck constant.
There exist a symplectic two-form
 denoted by $\sum_{i=1}^{d}dp_i\wedge dq^i$
 on the phase space.
For the quantization
 to be applicable to the system,
 integrals of the symplectic two-form
 over compact two-cycles in the phase space
 must be in $h \mathbb{Z}$.

Every $d$-dimensional toric variety is determined by 
 a fan \cite{toric}.  
The fan consists of cones
 in $M_{\mathbb{R}}$,
 where $M_{\mathbb{R}}= M \otimes \mathbb{R}$
 and $M$ is a $d$-dimensional lattice.
Each cone is generated by a finite number
 of vectors in the lattice,
 which are called primitive vectors,
 and 
 its apex is at the origin of $M$.
Let $e_i$ $(1\leq i \leq 3)$ be the generators
 of $M$.
If the toric variety is a Calabi-Yau manifold,
 the $e_3$ components
 of all the primitive vectors
 can be set to one.
We denote 
 such vectors
 by $v_{i\, j}$:
\begin{eqnarray}
v_{i\,j}= i e_1 + j e_2 + e_3,
\hspace{5mm}
i,j \in \mathbb{Z}.
\label{eq;v ij}
\end{eqnarray}

\begin{figure}[hbt]
\begin{center}
\psfrag{00}{\hspace{-3mm}$v_{0\,0}$}
\psfrag{03}{\hspace{-3mm}$v_{0\,3}$}
\psfrag{10}{$v_{1\,0}$}
\psfrag{-10}{$v_{-1\,0}$}
\includegraphics[scale=0.8]{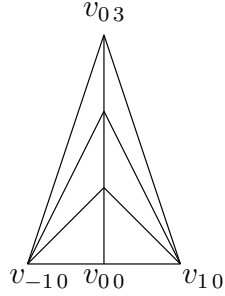}
 \caption{\textit{$\Sigma_{pure}$ 
 in the case of $N=3$ and $k_{\mbox{\tiny C.S.}}=N$.
}}
 \label{fig;sigma pure}
\end{center}
\end{figure}

The fan for the local $SU(N)$
 geometry consists of 
 the following three-dimensional cones
 besides their faces.
\begin{eqnarray}
\begin{array}{c}
\mathbb{R}_{\geq 0} v_{0\, j}
 + \mathbb{R}_{\geq 0} v_{0\, j+1}
 + \mathbb{R}_{\geq 0} v_{1\, 0},  \\
\mathbb{R}_{\geq 0} v_{0\, j}
 +\mathbb{R}_{\geq 0} v_{0\, j+1}
 + \mathbb{R}_{\geq 0} v_{-1\,  N-k_{\mbox{\tiny C.S.}} } 
\end{array}
\hspace{2mm}
\mbox{for all $0\leq j \leq N-1$},
\end{eqnarray}
where $v_{i\,j}$ are as above.
We denote the fan by $\Sigma_{pure}$
 and the set of
 indices of the primitive
 vectors by $S_{pure}$.
\begin{eqnarray}
S_{pure} = \left\{
(0, 0),\cdots,(0,N), (1,0),
 (-1, N-k_{\mbox{\tiny C.S.}} )
\right\}.
\end{eqnarray}
An example of $\Sigma_{pure}$ is shown in 
 Figure \ref{fig;sigma pure}.

We apply the Bohr-Sommerfeld quantization 
 to the local $SU(N)$ geometry
 by identifying the variety with the phase space,
 the K\"{a}hler two-form
 with the symplectic two-form
 and
 the string coupling $g_{st}$ with the Planck constant.
There are $N$ pieces of
 $\mathbb{CP}^1$ in the variety:
 One is the base $\mathbb{CP}^1$
 and the others are in the fiber.
The K\"{a}hler parameters
 for the base $\mathbb{CP}^1$
 and $\mathbb{CP}^1$ in the fiber
 are denoted by $t_B$ and $t_r$ respectively,
 where $r$ runs from $1$ to $N-1$. 
These parameters
 are quantized
 for the consistent quantization.
\begin{eqnarray}
T_r = t_r/ g_{st},\, T_B= t_B/ g_{st}
\hspace{1mm}
 \in \mathbb{Z}_{\geq 0}.
\end{eqnarray}
We further require the $SU(N)$ condition
 \cite{MNTT1}
 on the parameters.
Let us write $T_r$
 by using $N$ integers $p_r$
 as $T_r = p_{N-r+1}- p_{N-r}$.
We will impose the condition 
 $\sum_{r=1}^{N} p_{r}=0$.

The Hilbert space
 of the quantization
 can be read from
 a polyhedron.
Let $M^{\vee}$ the three-dimensional lattice 
 dual to $M$.
The dual pairing is denoted
 by $\langle , \rangle$.
Let $e_i^*$ $(1\leq i \leq 3)$
 be the generators of $M^{\vee}$,
 which satisfy 
 $\langle e_i^*, e_j \rangle =\delta_{ij}$.
The polyhedron
 $\mathcal{P}_{pure}$
 is defined as follows:
\begin{eqnarray}
 \mathcal{P}_{pure} = \left\{
 m \in M_{\mathbb{R}}^{\vee}| 
 \langle m , v_{i\, j} \rangle \geq -d_{i\, j},\, 
^\forall (i, j) \in S_{pure} 
 \right\}, \label{P pure}
\end{eqnarray}
where
 $d_{i\, j}$ are certain integers
 and $M_{\mathbb{R}}^{\vee}=M^{\vee} \otimes\mathbb{R}$.
The integers appear as the coefficients
 of a Weil divisor 
 $D=\sum_{(i,j)\in S_{pure}} d_{i\,j} D_{i\,j}$,
 where each $D_{i\,j}$ is
 an irreducible divisor corresponding to
 the edge $\mathbb{R}_{\geq 0} v_{i\,j}$.
The polyhedron can be regarded
 as 
 an image of the moment map
 of real 3$d$ torus actions.
A compact edge of the polyhedron
 corresponds to $\mathbb{CP}^1$
 in the geometry
 and its length corresponds to
 the K\"{a}hler parameter.
Therefore $d_{i\,j}$
 are chosen such that
 the following combinations
 give the K\"{a}hler parameters.
\begin{eqnarray}
T_r
&:=&
d_{0\, r-1}-2d_{0\, r}+ d_{0\, r+1} \geq 0, \nonumber \\
T_B
&:=&
-2d_{0\, 0} +d_{1\, 0}+d_{-1\, N-k_{\mbox{\tiny C.S.}}}
+\frac{N-k_{\mbox{\tiny C.S.}}}{N}(d_{0\, 0}-d_{0\, N}) \geq 0.
\label{eq;T_i T_B dij}
\end{eqnarray}

\begin{figure}[hbt]
\begin{center}
\psfrag{00}{\hspace{-3mm}$K_{0\,0}$}
\psfrag{01}{\hspace{-3mm}$K_{0\,1}$}
\psfrag{02}{\hspace{-3mm}$K_{0\,2}$}
\psfrag{03}{\hspace{-3mm}$K_{0\,3}$}
\psfrag{10}{$K_{1\,0}$}
\psfrag{-10}{$K_{-1\,0}$}
\includegraphics[scale=0.8]{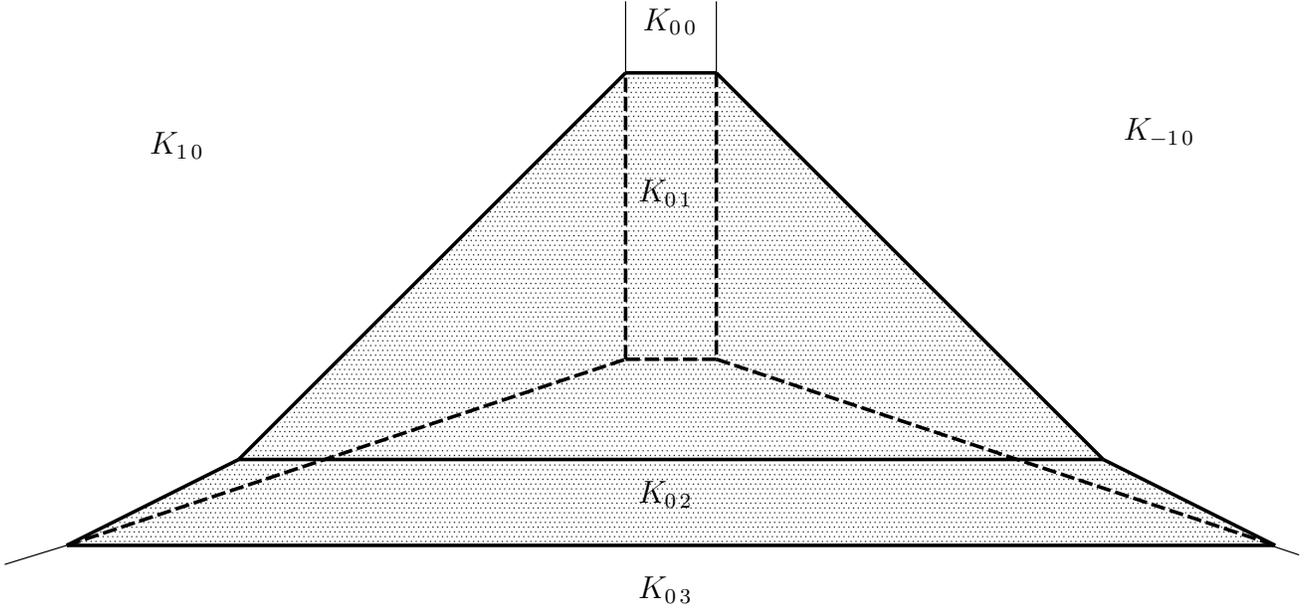}
 \caption{\textit{The shaded solid in the figure is
$\mathcal{P}_{pure}^c$
 in the case of $N=3$ and $k_{\mbox{\tiny C.S.}}=N$.
}}
 \label{fig;p^c_pure}
\end{center}
\end{figure}

It turns out that
 the base vectors of 
 the Hilbert space $\mathcal{H}_{pure}$
 are labelled by
 integer lattice points in 
 $\mathcal{P}_{pure} \cap M^{\vee}$
 and
 the dimension of $\mathcal{H}_{pure}$
 is
 the cardinality of 
 $\mathcal{P}_{pure}\cap M^{\vee}$.
The dimension 
 is infinite
 since
 the cardinality
 is infinite.
Nevertheless,
 we can regularize it \cite{MNNT}
 by considering a deviation from
 the singular limit
 where
 $T_r \rightarrow 0$
 with keeping $T_B$ fixed.
Let $\Theta_{pure}$ be the polyhedron
 $\mathcal{P}_{pure}$
 which appears at the limit.
It is surrounded by four planes:
 $K_{0\, 0}$,
 $K_{0\, N}$,
 $K_{1\, 0}$ and
 $K_{-1\, N-k_{\mbox{\tiny C.S.}}}$,
 where 
 each $K_{i\, j}$ is a plane
 and orthogonal to $v_{i\,j}$.
\begin{eqnarray}
 K_{i\, j} = \{
 m \in M_{\mathbb{R}}^{\vee}| 
 \langle m , v_{i\, j} \rangle = -d_{i\, j} 
 \}. \label{eq;K ij}
\end{eqnarray}
Let $\mathcal{P}_{pure}^c$
 be
 the complement of $\mathcal{P}_{pure}$
 in $\Theta_{pure}$.
\begin{eqnarray}
\mathcal{P}_{pure}^c 
= \mbox{Cl}(\Theta_{pure} \setminus\mathcal{P}_{pure}),
\end{eqnarray}
 where $\mbox{Cl}$ means 
 to take the closure.
An example of $\mathcal{P}_{pure}^c$
 is shown in Figure \ref{fig;p^c_pure}.
We define the dimension of $\mathcal{H}_{pure}$
 as the cardinality of 
 $\mathcal{P}_{pure}^c \cap M^{\vee}$:
\begin{eqnarray}
 g_{st}\cdot\mbox{dim}\, \mathcal{H}_{pure}
=
-g_{st} \mbox{Card}(\mathcal{P}_{pure}^c \cap M^{\vee}),
\end{eqnarray}
where the sign is a convention.

As $g_{st} \rightarrow 0$ with 
 keeping $t_r$ and $t_B$ fixed,
 the cardinality
 becomes
 the volume of $\mathcal{P}_{pure}^c$,
 which is denoted by
 $\mbox{Vol} (\mathcal{P}_{pure}^c)$.
It is calculated in \cite{MNNT} as follows:
\begin{eqnarray}
 \mbox{Vol} (\mathcal{P}_{pure}^c)
&=&
\sum_{n=1}^{N-1}
\frac{1}{3n(n+1)}
\left(
 \sum_{r=1}^{n} rT_r\right)^3  
+T_B
\sum_{n=1}^{N-1}\frac{1}{2n(n+1)} 
\left(
 \sum_{r=1}^{n} rT_r \right)^2 \nonumber \\
 &&
  +\frac{N-k_{\mbox{\tiny{C.S.}}}}{6}
  \sum\limits_{n=1}^{N}
  \left(
   \frac{1}{N}\sum\limits_{r=1}^{N-1}
   \left(N-r \right)T_r
   -\sum\limits_{r=1}^{k-1}T_r
  \right)^3 .
\end{eqnarray}
Let us make a relation between the geometric parameters
 and the gauge theoretic parameters
 as follows:
\begin{eqnarray}
g_{st}=\beta\hbar,\hspace{5mm}
\tilde{p}_{r}= \frac{a_r}{\hbar}, \hspace{5mm}
g_{st}T_B= -2N\ln (\beta\Lambda),
 \label{eq;id geo and gauge}
\end{eqnarray}
 where
 $\beta$ is the circumference of the circle in the fifth-direction,
 $\hbar$ is a parameter,
 $\tilde{p}_{r}=p_r+\xi_r$,
 $\xi_r=\frac{1}{N}\left(r -\frac{N+1}{2}\right)$,
 $a_{r}$ are vev. of the adjoint scalar and
 $\Lambda$ is the scale parameter
 for the underlying four-dimensional theory.
With this identification,
 the perturbative prepotential 
 for the five-dimensional $SU(N)$ gauge theory
 with no matters
 emerges from $\mbox{dim}\, \mathcal{H}_{pure}$
 as $\hbar\rightarrow 0$.
\begin{eqnarray}
-g_{st} \mbox{Vol} (\mathcal{P}_{pure}^c)
\stackrel{\hbar\rightarrow 0}{\longrightarrow}
-\frac{1}{\hbar^2}
\mathcal{F}^{pert}_{pure}(a_r;\beta, \ln (\beta\Lambda),
  k_{\mbox{\tiny{C.S.}}})
+ \mathcal{O}(\hbar^{-1})
,
\end{eqnarray}
where
\begin{eqnarray}
\mathcal{F}^{pert}_{pure}(a_r;\beta , \ln (\beta\Lambda),
  k_{\mbox{\tiny{C.S.}}})
=
\frac{\beta}{6}\sum_{r>s}^{N}a_{rs}^3
-\frac{\beta k_{\mbox{\tiny C.S.}}}{6}\sum_{r=1}^{N} a_r^3
-\ln (\beta\Lambda)\sum_{r>s}^N a_{rs}^2
.
\end{eqnarray}

\subsection{The case of $1\leq N_F\leq 2N$}

We describe the relation
 between the Hilbert space
 for a certain non-compact
 toric variety
 under
 the Bohr-Sommerfeld quantization
 and the perturbative prepotential for
 the five-dimensional $SU(N)$ gauge theory with
 $N_F$ ($1 \leq N_F \leq 2N$) 
 massive fundamental matters.
There are more $N_F$
 one-dimensional cones in the fan
 for the variety
 than the number of 
 one-dimensional cones 
 in $\Sigma_{pure}$.
As in the 
 previous section,
 the geometry is quantized
 and
 states in the Hilbert space
 turn to be labelled
 by
 integer lattice points
 in a certain polyhedron.
We can obtain the perturbative prepotential
 for the five-dimensional gauge theory
 with $N_F$ massive fundamental matters
 from the volume of the polyhedron.

\begin{figure}[hbt]
\begin{center}
\psfrag{00}{\hspace{-3mm}$v_{0\,0}$}
\psfrag{03}{\hspace{-3mm}$v_{0\,3}$}
\psfrag{10}{$v_{1\,0}$}
\psfrag{11}{$v_{1\,1}$}
\psfrag{12}{$v_{1\,2}$}
\psfrag{13}{$v_{1\,3}$}
\psfrag{-10}{$v_{-1\,0}$}
\psfrag{-11}{$v_{-1\,1}$}
\psfrag{01}{}
 \includegraphics{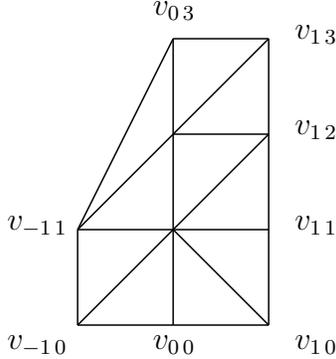}
 \caption{\textit{$\Sigma_{N_F}$ 
 in the case of  $N=3$, $N_L=3$,
$(l_1,l_2,l_3)=(1,1,2)$, $N_R=1$ and $r_1=1$.
}}
 \label{fig;Sigma N_F}
\end{center}
\end{figure}

We introduce 
 the following 
 integers $N_L$, $N_R$, $\{l_f\}$ and $\{r_f\}$:
\begin{eqnarray}
& N_L\in[1,N], \, N_R\in [0,N]\hspace{5mm} N_L+N_R=N_F \\
&0=l_0 < l_1 \leq \cdots \leq l_{N_L} < l_{N_{L}+1} =N, \hspace{5mm}
0=r_0 < r_1 \leq \cdots \leq r_{N_R}< r_{N_{R}+1}= N.
 \label{eq;l_f r_f}
\end{eqnarray}
The following
 three-dimensional cones 
 besides their faces
 provide a fan for the geometry.
\begin{eqnarray}
\begin{array}{cl}
\mathbb{R}_{\geq 0}v_{0\, j}
 +\mathbb{R}_{\geq 0}v_{0\, j+1}
 +\mathbb{R}_{\geq 0}v_{1\, f}
&
 \mbox{for $l_f\leq j\leq l_{f+1}-1$,
       $0\leq f \leq N_{L}$,} \\
\mathbb{R}_{\geq 0}v_{0\, l_f}
 +\mathbb{R}_{\geq 0}v_{1\, f-1}
 +\mathbb{R}_{\geq 0}v_{1\, f}
&
 \mbox{for $1\leq f \leq N_{L}$,} \\
 \mathbb{R}_{\geq 0}v_{0\, j}
 +\mathbb{R}_{\geq 0}v_{0\, j+1}
 +\mathbb{R}_{\geq 0}v_{-1\, f}
&
 \mbox{for $r_f\leq j\leq r_{f+1}-1$,
       $0\leq f \leq N_{R}$,} \\
\mathbb{R}_{\geq 0}v_{0\, r_f}
 +\mathbb{R}_{\geq 0}v_{-1\, f-1}
 +\mathbb{R}_{\geq 0}v_{-1\, f}
&
 \mbox{for $1\leq f \leq N_{R}$.}
\end{array}
\end{eqnarray}
We denote the fan by $\Sigma_{N_F}$ 
 and the set
 of indices of the primitive vectors
 by $S_{N_F}$.
\begin{eqnarray}
S_{N_F}=\left\{
(0, 0),\cdots,(0,N),
 (1,0),\cdots, (1,N_L),
 (-1,0),\cdots, (-1,N_R)
\right\}.
\end{eqnarray}
An example of $\Sigma_{N_F}$ is shown 
 in Figure \ref{fig;Sigma N_F}.
The other choices of $l_f,\, r_f$
 in (\ref{eq;l_f r_f})
 yield geometries
 which are related
 with each other  
 by flop transitions.

We apply the Bohr-Sommerfeld quantization
 to the geometry.
The base vectors of 
 the Hilbert space $\mathcal{H}_{N_F}$
 of the quantization
 is labelled by
 points in the intersection of
 $M^\vee$ and
 the following polyhedron
 $\mathcal{P}_{N_F}$:
\begin{eqnarray}
 \mathcal{P}_{N_F} = \left\{
 m \in M_{\mathbb{R}}^{\vee}| 
 \langle m , v_{i\, j} \rangle \geq -d_{i\, j},\, 
^\forall (i,j) \in S_{N_F}
 \right\},
\end{eqnarray}
 where 
 $d_{i\, j}$ are certain integers.
The integers appear as the coefficients
 of a Weil divisor 
 $D=\sum_{(i,j)\in S_{N_F}} d_{i\,j} D_{i\,j}$,
 where each $D_{i\,j}$ is
 an irreducible divisor corresponding to
 the edge $\mathbb{R}_{\geq 0} v_{i\,j}$.
Their linear combinations
 are identified with the K\"{a}hler parameters
 by using the moment map.
It turns out that
 the K\"{a}hler parameters are
 determined by
 $T_B$, $T_r$,
 which are given
 in (\ref{eq;T_i T_B dij}),
 and the following $N_F$ parameters $T_{m_f}$:
\begin{eqnarray}
T_{m_f}
=
 \begin{cases}
\frac{1}{N}(d_{0\,0}-d_{0\, N})
 -d_{1\,f-1}+d_{1\,f}& \mbox{for $1\leq f\leq N_L$}\\
\frac{1}{N}(d_{0\,0}-d_{0\, N})
 -d_{-1\,f-N_L-1}+d_{-1\,f-N_L}&
 \mbox{for $N_L +1\leq f\leq N_F$}.
\end{cases}
\end{eqnarray}
We require 
\begin{eqnarray}
T_B,
\hspace{1mm}
T_r,
\hspace{1mm}
T_{m_f} \geq 0,
\hspace{5mm}
^\forall r\in[1,N-1],\,\,
^\forall f\in[1,N_F].
\end{eqnarray}
We further require
 the $SU(N)$ condition \cite{MNTT1} on $T_r$
 and
 the following conditions\footnote{By this condition,
 the line bundle $\mathcal{O}(D)$
 is ample.}:
\begin{eqnarray}
\begin{array}{ccc}
T_{l_f}
\geq
&
\frac{1}{N}\sum_{r=1}^N
 (N-r)T_r
- \sum_{r=1}^{l_f-1}T_r
+T_{m_f}\geq 0
&
\mbox{for $1\leq f \leq N_L$},
 \\
T_{r_f}
\geq 
&
\frac{1}{N}\sum_{r=1}^N
 (N-r)T_r
- \sum_{r=1}^{r_f-1}T_r
+T_{m_{N_L+f}}\geq 0
&
\mbox{for $1\leq f \leq N_R$}.
\end{array} 
\label{eq;ample N_F}
\end{eqnarray}
$T_{m_{f}}$
 appear in $M^{\vee}_{\mathbb{R}}$
 as distances between intersection points of $K_{i\,j}$,
 where $K_{i\,j}$ are defined in (\ref{eq;K ij}).
\begin{eqnarray}
T_{m_{f}}
=
\begin{cases}
(K_{0\, 0}\cap K_{0\, N}\cap K_{1\,0})_{e_2^*}
-
(K_{0\, N}\cap K_{1\, f-1}\cap K_{1\,f})_{e_2^*} 
& \mbox{for $1\leq f\leq N_L$}, \\
(K_{0\, 0}\cap K_{0\, N}\cap K_{-1\,0})_{e_2^*}
-
(K_{0\, N}\cap K_{-1\, f-N_L-1}\cap K_{-1\,f-N_L})_{e_2^*} 
& \mbox{for $N_L+1\leq f\leq N_F$},
\end{cases}
\end{eqnarray}
where
 the subscript
 means to take
 the $e_i^*$ component of the points.

\begin{figure}[hbt]
\begin{center}
\psfrag{00}{\hspace{-3mm}$K_{0\,0}$}
\psfrag{01}{\hspace{-3mm}$K_{0\,1}$}
\psfrag{02}{\hspace{-3mm}$K_{0\,2}$}
\psfrag{03}{\hspace{-3mm}$K_{0\,3}$}
\psfrag{10}{$K_{1\,0}$}
\psfrag{11}{$K_{1\,1}$}
\psfrag{12}{$K_{1\,2}$}
\psfrag{13}{$K_{1\,3}$}
\psfrag{-10}{$K_{-1\,0}$}
\psfrag{-11}{$K_{-1\,1}$}
\psfrag{Tm1}{$T_{m_1}$}
\psfrag{Tm2}{$T_{m_2}$}
\psfrag{Tm3}{$T_{m_3}$}
\psfrag{Tm4}{$T_{m_4}$}
 \includegraphics[scale=0.7]{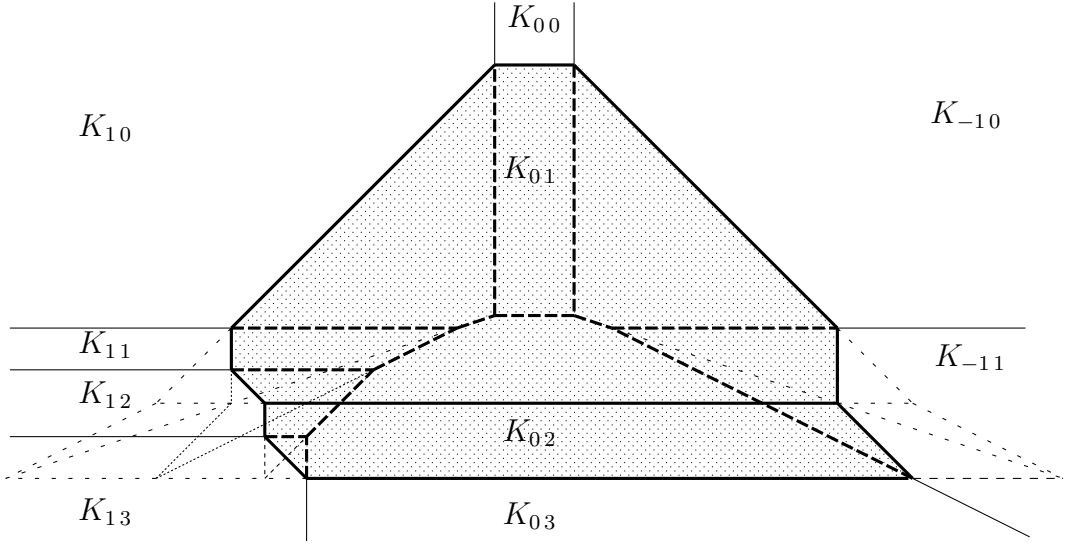}
 \caption{\textit{The shaded solid in the figure is
$\mathcal{P}_{N_F}^c$ 
in the case of $N=3$, $N_L=3$,
$(l_1,l_2,l_3)=(1,1,2)$, $N_R=1$ and $r_1=1$.}}
 \label{fig;P^c_N_F}
\end{center}
\end{figure}

Let $\Theta_{N_F}$ be the polyhedron
 $\mathcal{P}_{N_F}$
 which appears
 at the limit
 where
 $T_r\rightarrow 0$
 with keeping $T_B$ and $T_{m_{f}}$ fixed.
$\Theta_{N_F}$ is surrounded by
 $N+N_F+3$ planes:
 $K_{0\,0}$, $K_{0\,N}$, 
 $K_{1\,0}$,$\cdots$, $K_{1\,N_L}$, 
 $K_{-1\,0}$, $\cdots$, $K_{-1\, N_R}$. 
Each apex of $\Theta_{N_F}$ is on $M^{\vee}$
 due to the $SU(N)$ condition.
Let $\mathcal{P}_{N_F}^c$
 be
 the complement of
 $\mathcal{P}_{N_F}$ in $\Theta_{N_F}$.
\begin{eqnarray}
\mathcal{P}_{N_F}^c
=
\mbox{Cl}(
\Theta_{N_F} \setminus\mathcal{P}_{N_F}),
\end{eqnarray}
where $\mbox{Cl}$ means to take the closure.
An example of $\mathcal{P}_{N_F}^c$ 
 is shown in Figure \ref{fig;P^c_N_F}.
We regularize  
 the dimension
 of the Hilbert space
 by using 
 $\mathcal{P}_{N_F}^c$.
\begin{eqnarray}
 g_{st}\cdot\mbox{dim}\, \mathcal{H}_{N_F}
=
-g_{st} \mbox{Card}(\mathcal{P}_{N_F}^c \cap M^{\vee}).
\end{eqnarray}

\begin{figure}[hbt]
\hspace{-5mm}
\begin{minipage}{170mm}
\psfrag{01}{$K_{0\,1}$}
\psfrag{02}{$K_{0\,2}$}
\psfrag{03}{$K_{0\,3}$}
\psfrag{10}{$K_{1\,0}$}
\psfrag{11}{$K_{1\,1}$}
\psfrag{12}{\hspace{-1mm}$K_{1\,2}$}
\psfrag{13}{$K_{1\,3}$}
 \includegraphics[scale=1.0]{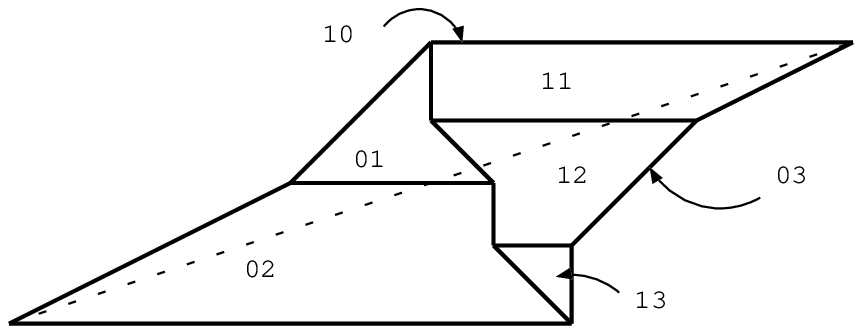}
\hspace{5mm}
\psfrag{-10}{$K_{-1\,0}$}
\psfrag{-11}{$K_{-1\,1}$}
 \includegraphics[scale=1.0]{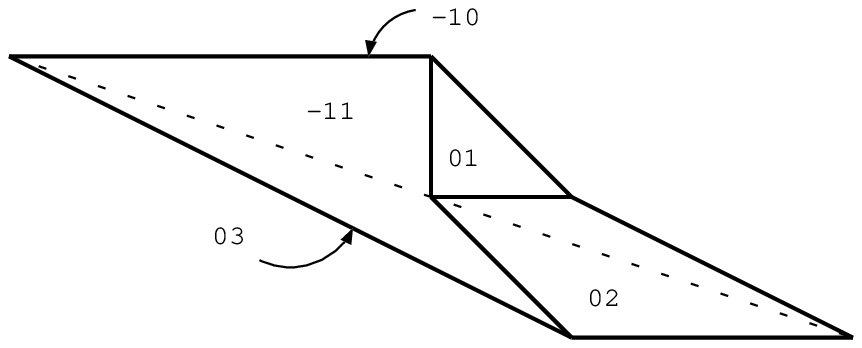}
 \caption{\textit{ $P_L$ $(left)$ and  $P_R$ $(right)$
in the case of $N=3$, $N_L=3$,
$(l_1,l_2,l_3)=(1,1,2)$, $N_R=1$ and $r_1=1$.}}
 \label{fig;P_l P_r}
\end{minipage}
\end{figure}

\begin{figure}[hbt]
\begin{center}
\psfrag{Pp}{$\mathcal{P}_{pure}^c$}
\psfrag{PN}{$\mathcal{P}_{N_F}^c$}
\psfrag{PL}{$P_L$}
\psfrag{PR}{$P_R$}
 \includegraphics[scale=0.7]{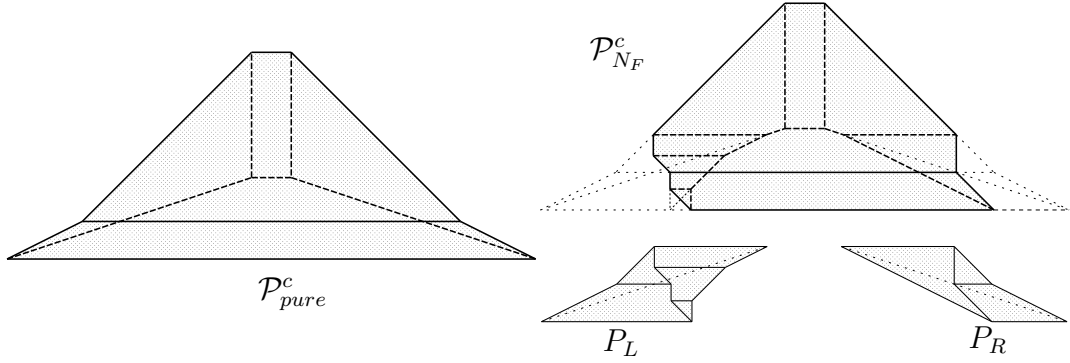}
 \caption{\textit{
The relation of 
$\mathcal{P}^c_{pure}$,
$\mathcal{P}_{N_F}^c$,
 $P_L$ and $P_R$
 in the case of $N=3$, $N_L=3$,
 $(l_1,l_2,l_3)=(1,1,2)$, $N_R=1$ and $r_1=1$.}}
 \label{fig;removing}
\end{center}
\end{figure}

As $g_{st}\rightarrow 0$
 with keeping $g_{st}T_B$, $g_{st}T_r$ and $g_{st}T_{m_f}$
 fixed,
 $\mbox{dim}\, \mathcal{H}_{N_F}$
 approximates to
 the volume of $\mathcal{P}_{N_F}^c$ times minus one. 
As a solid,
 $\mathcal{P}_{N_F}^c$ is
 obtained
 from $\mathcal{P}^c_{pure}$
 in the case of $k_{\mbox{\tiny C.S.}}=N$
 by removing
 two solid 
 $P_L$ and $P_R$.
\begin{eqnarray}
\mathcal{P}_{N_F}^c
=\mathcal{P}_{pure}^c|_{k_{\mbox{\tiny C.S.}}=N}
 \setminus (P_L\cup P_R)
\end{eqnarray}
An example of
 $P_L$ and $P_R$ are shown in Figure \ref{fig;P_l P_r}
 and 
 the relation of 
 $\mathcal{P}^c_{pure}$,
 $\mathcal{P}_{N_F}^c$,
 $P_L$ and $P_R$
 is shown in Figure \ref{fig;removing}. 
We see below that
 the matter terms in the perturbative prepotential
 emerge 
 from
 the volume of $P_L$ and $P_R$.

\begin{figure}[hbt]
\begin{center}
\psfrag{0i}{$K_{0\,i}$}
\psfrag{0i+1}{$K_{0\,i+1}$}
\psfrag{1j}{$K_{1\,j}$}
\psfrag{1j+1}{$K_{1\,j+1}$}
\psfrag{T}{}
\psfrag{(a)}{$(a)$}
\psfrag{(b)}{$(b)$}
 \includegraphics[scale=0.7]{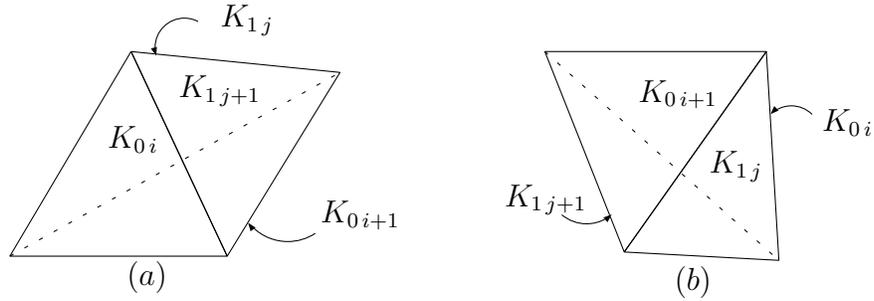}
 \caption{\textit{$P_{i\,j}$
 in the case of $(a)$ 
 $-d_{0\, i}+d_{0\, i+1} + d_{1\, j}-d_{1\, j+1} \geq 0$,
and
 $(b)$ $-d_{0\, i}+d_{0\, i+1} + d_{1\, j}-d_{1\, j+1} \leq 0$.}}
 \label{fig;P ij}
\end{center}
\end{figure}

We calculate the volume of $P_L$ and $P_R$. 
$P_L$ is sliced into
 triangular pyramids
 by 
 $K_{1\,n}$ ($1\leq n\leq N_L-1$)
 and $K_{0\,m}$ ($l_1+1\leq m\leq N-1$):
\begin{eqnarray}
P_L=
\bigcup_{j=1}^{N_L}
\bigcup_{i=l_j}^{N-1}
 P_{i\,j-1},
 \label{eq;Pij in P_L}
\end{eqnarray}
where
 each $P_{i\,j}$ is a triangular pyramid
 surrounded by four planes 
 $K_{0\, i}$, 
 $K_{0\, i+1}$,
 $K_{1\, j}$ and
 $K_{1\, j+1}$.
The pyramids are classified 
 into two types
 as 
 in Figure \ref{fig;P ij}
 according as
 $T=-d_{0\, i}+d_{0\, i+1} + d_{1\, j}-d_{1\, j+1}$
 takes a positive or negative value.
The pyramids in (\ref{eq;Pij in P_L})
 belong to type $(a)$ in Figure \ref{fig;P ij}.
The volume of such pyramid $P_{i\, j}$
 is $\frac{1}{6} T^3$.
Then the volume of $P_L$ becomes
\begin{eqnarray}
\mbox{Vol}(P_L)
&=&
 \sum_{j=1}^{N_L} \sum_{i=l_j}^{N-1}
\frac{1}{6}
\left(
 \sum_{r=1}^{i}T_r
 -\frac{1}{N}\sum_{r=1}^{N-1}(N-r)T_r
 -T_{m_{j}}
\right)^3,
\end{eqnarray}
where we have used the identities 
\begin{eqnarray}
&-d_{0\, i}+d_{0\, i+1} + d_{1\, j-1}-d_{1\, j}
=
 \sum_{r=1}^{i}T_r
 -\frac{1}{N}\sum_{r=1}^{N-1}(N-r)T_r
 -T_{m_{j}} \nonumber\\
&
\mbox{\hspace{60mm}for $l_j\leq i \leq N-1$,
 $1\leq j \leq N_L$} 
\end{eqnarray}
Similarly,
 $P_R$ is sliced into triangular pyramids
 and its volume becomes
\begin{eqnarray}
\mbox{Vol}(P_R)
&=&
 \sum_{j=1}^{N_R} \sum_{i=r_j}^{N-1}
\frac{1}{6}
\left(
 \sum_{r=1}^{i}T_r
 -\frac{1}{N}\sum_{r=1}^{N-1}(N-r)T_r
 -T_{m_{N_L+j}}
\right)^3.
\end{eqnarray}

Let us make a relation between the geometric parameters
 and the gauge theoretic parameters as follows:
\begin{eqnarray}
g_{st}=\beta\hbar,\hspace{5mm}
\tilde{p}_{r}= \frac{a_r}{\hbar}, \hspace{5mm}
g_{st}T_B
  = \frac{\beta}{g_{YM}^2}
   -\frac{\beta}{2}\sum_{f=1}^{N_F} m_f, \hspace{5mm}
T_{m_{f}} = \frac{m_f}{\hbar},\nonumber\\
\frac{\beta}{g_{YM}^2}=\begin{cases}
     -(2N-N_F)\ln(\beta \Lambda)  &
     \mbox{if \,$1\leq N_F < 2N$}\\
     const. & \mbox{if \,$N_F=2N$}
    \end{cases},
\end{eqnarray}
where $m_{f}$ are the mass of the fundamental matters
 and the meanings of the other parameters
 are same as (\ref{eq;id geo and gauge}).
With this identification we have
\begin{eqnarray}
-g_{st} \mbox{Card}(\mathcal{P}_{N_F}^c \cap M^{\vee})
&\stackrel{\hbar\rightarrow 0}{\longrightarrow}&
-g_{st}\mbox{Vol}(\mathcal{P}_{N_F}^c) \nonumber\\
&=&
-g_{st}\mbox{Vol}(\mathcal{P}_{pure}^c|_{k_{\mbox{\tiny C.S.}}=N})
+g_{st}\mbox{Vol}(P_L)
+g_{st}\mbox{Vol}(P_R) \nonumber \\
&=&
-\frac{1}{\hbar^2}
\mathcal{F}^{pert}_{N_F}(a_r;\beta ,\frac{\beta}{g_{YM}^2},
  N, m_f)
 -\frac{N\beta}{12 \hbar^2}
   \sum_{f=1}^{N_F} m_f^3
+\mathcal{O}(\hbar^{-1}),
  \label{eq;pre_N_F}
\end{eqnarray}
where
\begin{eqnarray}
\mathcal{F}^{pert}_{N_F}(a_r;\beta , \frac{\beta}{g_{YM}^2},
  k_{\mbox{\tiny{C.S.}}}, m_f)
&=&
\frac{\beta}{6}\sum_{r>s}^{N}a_{rs}^3
-\frac{\beta}{12}\sum_{r=1}^{N}|a_{rs}+m_f|^3
-\frac{\beta(2 k_{\mbox{\tiny C.S.}} -N_F)}{12}
 \sum_{r=1}^{N} a_r^3 
\nonumber\\
&&
 +\frac{\beta}{g_{YM}^2}
\frac{1}{2}\sum_{r=1}^N a_{r}^2
.
\end{eqnarray}

\section{Perturbative prepotentials with one adjoint matter
 from polyhedrons}
\label{sec;dim count adj}

In this section
 we derive the relation 
 between the Hilbert space
 for a certain non-compact geometry
 with a periodic boundary condition
 and the perturbative prepotential for
 the five-dimensional $SU(N)$ gauge theory with
 one massive adjoint matter.
The non-compact geometry
 is obtained from a toric variety
 with certain conditions.
As in the previous sections,
 the geometry is quantized
 and states in the Hilbert space
 of the quantization
 are labelled
 by
 integer lattice points
 in a certain polyhedron.
The periodic boundary condition
 is archived by imposing a
 periodic boundary condition on 
 the polyhedron.
This allows us to 
 consider only
 a fundamental region of it.
The perturbative prepotential
 for the five-dimensional gauge theory
 with one massive adjoint matter
 can be obtained
 from the volume 
 of the fundamental region.

\begin{figure}[hbt]
\begin{center}
\psfrag{00}{$v_{0\,0}$}
\psfrag{03}{$v_{0\,3}$}
\psfrag{10}{$v_{1\,0}$}
\psfrag{13}{$v_{1\,3}$}
\psfrag{-10}{$v_{-1\,0}$}
\psfrag{-13}{$v_{-1\,3}$}
 \includegraphics{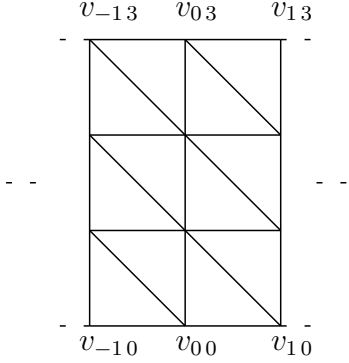}
 \caption{\textit{$\Sigma_{adj}$
 in the case of $N=3$.}}
 \label{fig;Sigma adj}
\end{center}
\end{figure}

First of all, 
 forgetting 
 the periodic boundary condition,
 we consider the toric variety.
The fan for the variety
 consists of the following
 three-dimensional cones
 besides their faces:
\begin{eqnarray}
\begin{array}{c}
\mathbb{R}_{\geq 0}v_{i\, j-1}
 +\mathbb{R}_{\geq 0}v_{i\, j}
 +\mathbb{R}_{\geq 0}v_{i+1\, j-1},
 \\
\mathbb{R}_{\geq 0}v_{i\, j}
 +\mathbb{R}_{\geq 0}v_{i+1\, j-1}
 +\mathbb{R}_{\geq 0}v_{i+1\, j} ,
\end{array}
\hspace{5mm}
^\forall i\in \mathbb{Z},\, ^\forall j\in [1,N].
\end{eqnarray}
We denote the fan by $\Sigma_{adj}$.
An example of $\Sigma_{adj}$
 are shown in Figure
 \ref{fig;Sigma adj}.

\begin{figure}[hbt]
\begin{center}
\psfrag{00}{$K_{0\,0}$}
\psfrag{01}{$K_{0\,1}$}
\psfrag{02}{$K_{0\,2}$}
\psfrag{03}{$K_{0\,3}$}
\psfrag{10}{\hspace{-3mm}$K_{1\,0}$}
\psfrag{11}{\hspace{-3mm}$K_{1\,1}$}
\psfrag{12}{\hspace{-3mm}$K_{1\,2}$}
\psfrag{13}{\hspace{-3mm}$K_{1\,3}$}
\psfrag{-10}{$K_{-1\,0}$}
\psfrag{-11}{$K_{-1\,1}$}
\psfrag{-12}{$K_{-1\,2}$}
\psfrag{-13}{$K_{-1\,3}$}
\psfrag{TB0}{$T_{B}$}
\psfrag{NTm0}{$NT_{m}$}
 \includegraphics[scale=0.7]{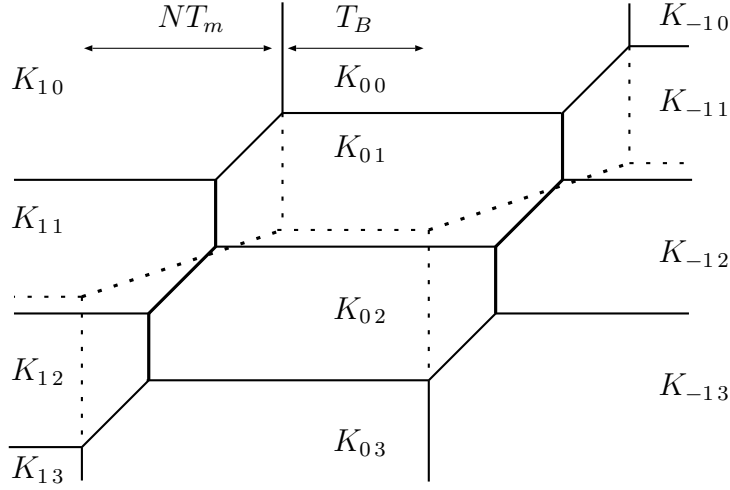}
 \caption{\textit{$\mathcal{P}_{adj}$
 in the case of $N=3$.}}
 \label{fig;P adj}
\end{center}
\end{figure}

We apply the Bohr-Sommerfeld
 quantization 
 to the geometry.
The Hilbert space $\mathcal{H}_{adj}$
 of the 
 quantization
 can be read from the following 
 polyhedron $\mathcal{P}_{adj}$:
\begin{eqnarray}
\mathcal{P}_{adj}
= \left\{
 m \in M_{\mathbb{R}}^{\vee}| 
 \langle m , v_{i\, j} \rangle \geq -d_{i\, j},\, 
  ^\forall i \in \mathbb{Z}, \,^\forall j \in[0,N]
 \right\},
\end{eqnarray}
 where 
 $d_{i\, j}$ are certain integers.
The integers appear as the coefficients
 of a Weil divisor 
 $D=\sum_{i \in \mathbb{Z}, \, 0\leq j \leq N}
 d_{i\,j} D_{i\,j}$,
 where each $D_{i\,j}$ is
 an irreducible divisor corresponding to
 the edge $\mathbb{R}_{\geq 0} v_{i\,j}$.
Their linear combinations
 are identified
 with the K\"{a}hler parameters
 by using the moment map.
It turns out that
 the K\"{a}hler parameters
 are determined by the following parameters:
\begin{eqnarray}
T_{B\,i}
=&
d_{i-1\, N}-d_{i\, 0}-d_{i\, N}+d_{i+1\, 0},
&
^\forall i \in \mathbb{Z}
\nonumber\\
T_{r\, i}
=&
d_{i\,r-1}-2d_{i\, r}+ d_{i\,r+1},
&
^\forall r \in[1,N-1],\,\,
^\forall i \in \mathbb{Z}
\label{eq;T_ni dij} \\
T_{m\, i}
=&
\frac{1}{N}(d_{i\, 0}-d_{i\, N}-d_{i+1\,0}+ d_{i+1\, N})
&
^\forall i \in \mathbb{Z}.
\nonumber
\end{eqnarray}
We require the parameters
 are 
 non-negative
 and
 independent of the subscript $i\in\mathbb{Z}$. 
\begin{eqnarray}
T_{B}=T_{B\, i}\geq 0,
\hspace{2mm}
T_r=T_{r\, i}\geq 0,
\hspace{2mm}
T_{m}=T_{m\, i}\geq 0,
\hspace{3mm}
^\forall r \in [1,N-1],\, ^\forall i \in \mathbb{Z}.
\label{eq;periodic}
\end{eqnarray}
We further require the following
 conditions\footnote{By this condition,
 the line bundle $\mathcal{O}(D)$ is ample.}:
\begin{eqnarray}
T_{s+1\,i}
\geq
&
\sum_{r=1}^{s}(T_{r\,i}-T_{r\,i+1})
-\frac{1}{N}\sum_{r=1}^{N-1}
 (N-r)(T_{r\,i}-T_{r\,i+1})
-T_{m\,i}
&
\geq 0,
\hspace{5mm}
 ^\forall s\in[1,N-2],\nonumber\\
&
\frac{1}{N}\sum_{r=1}^{N-1}
 (T_{r\,i}-T_{r\,i+1})
-T_{m\,i}
\geq 0.&
 \label{eq;combi adj}
\end{eqnarray}
An example of $\mathcal{P}_{adj}$
 is shown in Figure \ref{fig;P adj}.

By the condition (\ref{eq;periodic}),
 $\mathcal{P}_{adj}$
 can be seen as periodic 
 with periodicity $T_B+N T_m$
 along the $e_1^*$ direction.
The fundamental region
 $\widetilde{\mathcal{P}}_{adj}$
 can be taken 
 as follows:
\begin{eqnarray}
\widetilde{\mathcal{P}}_{adj}
= \left\{
 m \in M_{\mathbb{R}}^{\vee}
\left| 
\begin{array}{c}
 \langle m , v_{i\, j} \rangle \geq -d_{i\, j},\,\, 
^\forall i \in [0,1], \, ^\forall j\in[0,N], \\
 -T_B/2-NT_m
 <
 \langle m ,e_1 \rangle
 -(K_{0\,0}\cap K_{1\,0})_{e_1^*}
 \leq 
T_B/2
\end{array}
\right.
 \right\}.
\end{eqnarray}

Integer lattice points in
 $\widetilde{\mathcal{P}}_{adj}\cap M^{\vee}$
 label the base vectors
 of the Hilbert space $\mathcal{H}_{adj}$
 and
 the cardinality
 of
 $\widetilde{\mathcal{P}}_{adj}\cap M^{\vee}$
 is
 the dimension of the Hilbert space.
The dimension
 can be regularized
 by considering a deviation
 from
 the singular limit
 where
 $T_r\rightarrow 0$
 with keeping $T_B$ and $T_m$ fixed.
Let $\widetilde{\Theta}_{adj}$
 be the polyhedron $\widetilde{\mathcal{P}}_{adj}$
 which appears
 at the limit.
Each apex of $\widetilde{\Theta}_{adj}$
 is
 on $M^{\vee}$
 due to the $SU(N)$ condition.
Let $\widetilde{\mathcal{P}}_{adj}^c$
 be the complement
 of  $\widetilde{\mathcal{P}}_{adj}$ in 
 $\widetilde{\Theta}_{adj}$.
\begin{eqnarray}
 \widetilde{\mathcal{P}}_{adj}^c
=
\mbox{Cl}(
 \widetilde{\Theta}_{adj} \setminus
 \widetilde{\mathcal{P}}_{adj}),
\end{eqnarray}
 where $\mbox{Cl}$ means
 to take the closure.
The dimension of the Hilbert space 
 is defined as
 the cardinality of
 $\widetilde{\mathcal{P}}_{adj}^c \cap M^{\vee}$.
\begin{eqnarray}
g_{st}\cdot\mbox{dim}\, \mathcal{H}_{adj}
 =-g_{st}\mbox{Card}(\widetilde{\mathcal{P}}_{adj}
\cap M^\vee).
\end{eqnarray}

\begin{figure}[hbt]
\begin{center}
\psfrag{00}{$K_{0\,0}$}
\psfrag{01}{$K_{0\,1}$}
\psfrag{02}{$K_{0\,2}$}
\psfrag{03}{$K_{0\,3}$}
\psfrag{10}{$K_{1\,0}$}
\psfrag{11}{$K_{1\,1}$}
\psfrag{12}{$K_{1\,2}$}
\psfrag{13}{$K_{1\,3}$}
 \includegraphics[scale=0.7]{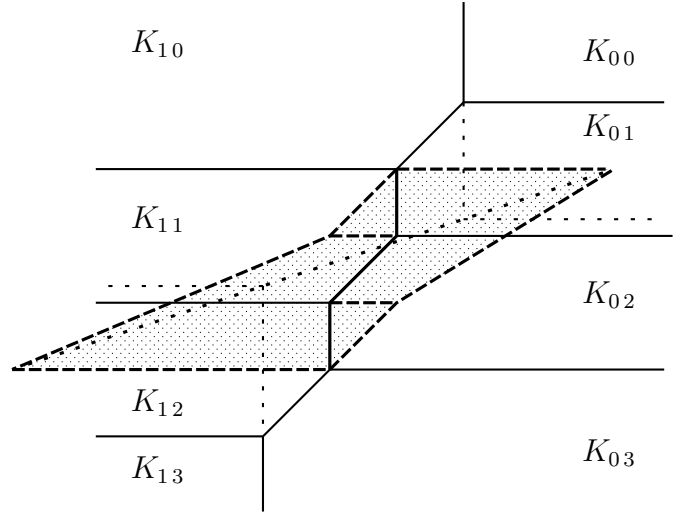}
 \caption{\textit{The shaded solid in the figure is
$P_{adj}$
in the case of $N=3$.}}
 \label{fig;P^c_adj}
\end{center}
\end{figure}

As $g_{st}\rightarrow 0$,
 the dimension 
 approximates to
 the volume of $\widetilde{\mathcal{P}}_{adj}^c$.
As a solid,
 $\widetilde{\mathcal{P}}_{adj}^c$
 is obtained by using certain
 two solids $P_0$ and $P_1$:
\begin{eqnarray}
\widetilde{\mathcal{P}}_{adj}^c
= (P_0\cup P_1)\cap \widetilde{\Theta}_{adj},
\end{eqnarray}
\begin{eqnarray}
P_0
&=& \left\{
  \sum_{i=1}^3 m_ie_i^* \in M_{\mathbb{R}}^{\vee}
\left|
\begin{array}{c}
\displaystyle
\sum_{i=1}^3 m_ie_i^* \in\mathcal{P}_{pure}^c|_{k_{\mbox{\tiny C.S.}}=0},
\hspace{2mm}
 m_1 
 -(K_{0\,0}\cap K_{1\,0})_{e_1^*}
 \leq
T_B/2
\end{array}
\right.
 \right\},\\
P_1
&=&
 \left\{
 \sum_{i=1}^3 m_ie_i^* \in M_{\mathbb{R}}^{\vee}
\left|
\begin{array}{l}
(m_1+T_B+NT_m)e_1^*
+(m_2+T_m)e_2^* \\
\,\,+(m_1+m_3-(K_{0\,0}\cap K_{1\,0})_{e_1^*})e_3^*
\in\mathcal{P}_{pure}^c|_{k_{\mbox{\tiny C.S.}}=0},\\
m_1-(K_{0\,0}\cap K_{1\,0})_{e_1^*}
> -T_B/2-NT_m
\end{array}
 \right.
 \right\}.
\end{eqnarray}
The sum of the 
 volume of $P_0$ and $P_1$
 is equal to $\mbox{Vol}(\mathcal{P}_{pure}^c|_{k_{\mbox{\tiny C.S.}}=0})$.
\begin{eqnarray}
\mbox{Vol}(P_0)+
\mbox{Vol}(P_1)=
\mbox{Vol}(\mathcal{P}_{pure}^c|_{k_{\mbox{\tiny C.S.}}=0}).
\end{eqnarray}
Then the volume of $\widetilde{\mathcal{P}}_{adj}^c$
 is
\begin{eqnarray}
\mbox{Vol}(\widetilde{\mathcal{P}}_{adj}^c)
=
\mbox{Vol}(\mathcal{P}_{pure}^c|_{k_{\mbox{\tiny C.S.}}=0})
-\mbox{Vol}(P_{adj}),
\end{eqnarray}
where $P_{adj}$ is 
\begin{eqnarray}
P_{adj}=(P_0\cap P_1)
\cup((P_0\cup P_1)
\setminus((P_0\cup P_1)\cap\widetilde{\Theta}_{adj})).
\end{eqnarray}
An example of $P_{adj}$ 
 is shown in Figure \ref{fig;P^c_adj}.

As a solid,
 $P_{adj}$
 is $P_L$, which appear
 in section \ref{sec;dim count fun},
 in the case of $N_L=N-1$ 
 and $l_f=f$ for $1\leq f \leq N-1$.
Hence $P_{adj}$
 consists of the triangular pyramids 
 $P_{i\,j}$.
\begin{eqnarray}
P_{adj}=
\bigcup_{j=0}^{N-2}
\bigcup_{i=j+1}^{N-1}
 P_{i\,j}.
\end{eqnarray}
By using equalities 
 $-d_{0\, i}+d_{0\, i+1} + d_{1\, j}-d_{1\, j+1}=
 \sum_{r=j+1}^{i}T_r -T_m$ for 
 $0\leq j<i \leq N-1$,
 we obtain
\begin{eqnarray}
\mbox{Vol}(P_{adj})
&=&
\frac{1}{6}
\sum_{j=0}^{N-2}\sum_{i=j+1}^{N-1}
\left(
\sum_{r=j+1}^{i}T_r -T_m
\right)^3.
\end{eqnarray}

Let us
 make an identification between 
 the
 geometric parameters and the gauge theory parameters
 as follows:
\begin{eqnarray}
&
g_{st}=\beta\hbar,\hspace{5mm}
T_r=p_{N-r+1}-p_{N-r},\hspace{5mm}
\tilde{p}_{r}= \frac{a_r}{\hbar}, \hspace{5mm}
T_{m} = \frac{m}{\hbar}, &\nonumber\\
&
g_{st}T_B= - \ln q -N \beta m,\hspace{5mm}
q=\exp(2\pi i \tau), \hspace{5mm}
\tau= \frac{4\pi i}{g_{YM}^2},
&
\end{eqnarray}
where $m$ is the mass for the adjoint matter,
 $1/g_{YM}^2$ is  
 the gauge coupling constant 
 and the meanings of the other parameters
 are same as (\ref{eq;id geo and gauge}).
With this identification,
 we can obtain
 the perturbative prepotential
 for the five-dimensional gauge theory 
 with one massive adjoint matter
 in the case of $a_{r+1}-a_r > m$
 from the volume of $\widetilde{\mathcal{P}}_{adj}^c$.
\begin{eqnarray}
 -g_{st}\mbox{Vol}(\widetilde{\mathcal{P}}_{adj}^c)
&=&
 -g_{st}\mbox{Vol}(\mathcal{P}_{pure}^c|_{k_{\mbox{\tiny C.S.}}=0})
 +g_{st}\mbox{Vol}(P_{adj}) \\
&\stackrel{\hbar \rightarrow 0}{\longrightarrow}&
 -\frac{1}{\hbar^2}
\mathcal{F}^{pert}_{adj}(a_r;\beta , - \ln q,
  0, m)
 -\frac{\beta N(N-1)m^3}{12\hbar^2}
+\mathcal{O}(\hbar^{-1})
,
\end{eqnarray}
where
\begin{eqnarray}
\mathcal{F}^{pert}_{adj}(a_r;\beta ,- \ln q ,
  k_{\mbox{\tiny{C.S.}}}, m)
&=&
\frac{\beta}{6}\sum_{r>s}^{N}a_{rs}^3
-\frac{\beta}{12}\sum_{r>s}^{N}(a_{rs}-m)^3
-\frac{\beta}{12}\sum_{r>s}^{N}(a_{rs}+m)^3
-\frac{\beta k_{\mbox{\tiny C.S.}}}{6}\sum_{r=1}^{N} a_r^3 
\nonumber\\
&&
 -
\frac{\ln q}{2N}\sum_{r>s}^N a_{rs}^2
. \label{eq;pre pert adj}
\end{eqnarray}

\section{Several features of
 the $\mathcal{N}=1^*$ gauge theory}
\label{sec;cft and N=1*}

The five-dimensional
 gauge theories with one massive 
 adjoint matter,
 named
 the $\mathcal{N}=1^*$ gauge theories,
 have interesting features.
We find that
 a five-dimensional generalization
 of Nekrasov's partition function
 for $U(1)$ gauge theory
 can be
 written as
 a correlation function
 of two-dimensional free chiral bosons.
This property
 of Nekrasov's partition function
 for the four-dimensional 
 gauge theory is described in 
 \cite{nek okoun}.
We investigate
 the partition function
 for the five-dimensional gauge theory
 by taking 
 several limits of the parameters.
By embedding $N$ partitions into 
 a single partition,
 we can also obtain
 a five-dimensional generalization
 of Nekrasov's partition function
 for the $SU(N)$ gauge theory
 from that
 for $U(1)$ gauge theory.

\subsection{A $5$D generalization
 of Nekrasov's partition function 
and $2$D CFT}

We describe here the relation between 
 a five-dimensional generalization
 of Nekrasov's partition function
 and a correlation function
 of two-dimensional free chiral bosons.
Let
\begin{eqnarray}
\mu=\frac{m}{\hbar}\in \mathbb{Z}_{\geq 0},
\hspace{5mm}
t=e^{-\beta\hbar},
\end{eqnarray}
 where $m$ is the mass of the adjoint matter.
The following equation will be 
 a five-dimensional generalization
 of Nekrasov's partition function
 for $U(1)$ gauge theory.
\begin{eqnarray}
Z_{\mathcal{N}=1^*}(q,t,\mu)
&=&
\sum_{\lambda} q^{|\lambda |}
\prod_{i\not= j}
\frac{
 \left[
  \lambda_i-i -\lambda_j +j
 \right]_{t^{1/2}}
}{
 \left[
  j-i
 \right]_{t^{1/2}}
}
\frac{
 \left[
 \mu+ j-i
 \right]_{t^{1/2}}
}{
 \left[
 \mu+ \lambda_i-i -\lambda_j +j
 \right]_{t^{1/2}}
} ,\label{eq;nek formula}
\end{eqnarray}
where
 $\lambda=(\lambda_1,\lambda_2,\cdots)$
 is a partition,
 which is a non-increasing sequence
 of non-negative integers,
 and
 $|\lambda|=\sum_{i=1}^{\infty}\lambda_i$.
In the above equation,
 we have used
\begin{eqnarray}
[n]_{t^{1/2}}=\frac{t^{n/2}-t^{-n/2}}{t^{1/2}-t^{-1/2}},
\end{eqnarray}
 which is called a ``$t^{1/2}$-integer''.
This partition function
 is symmetric under
 $t \leftrightarrow t^{-1}$.
\begin{eqnarray}
Z_{\mathcal{N}=1^*}(q,1/t,\mu)
= Z_{\mathcal{N}=1^*}(q,t,\mu)
\end{eqnarray}

A two-dimensional chiral boson $\varphi$
 is introduced as follows:
\begin{eqnarray}
\varphi(z)= -i J_0 \ln z 
+i \sum_{\substack{n=-\infty\\
                   n\not= 0}}^{\infty}
		    \frac{1}{n}\frac{J_n}{ z^n}, 
\hspace{5mm}
[J_k,J_{k'}]=k\delta_{k+k',0}.
\end{eqnarray}
We will see below
 that
 $Z_{\mathcal{N}=1^*}(q,t,\mu)$
 is expressed 
 by using $\varphi(z)$ as follows:
\begin{eqnarray}
Z_{\mathcal{N}=1^*}(q,t,\mu)
&=&
\mbox{Tr}\left(
q^{L_0}:\prod_{n=1}^{\mu}
\exp(-i\varphi(t^{-n+\frac{\mu+1}{2}})):
\right),  \label{eq;Z and correlation}
\end{eqnarray}
 where
 $L_0$ is the zero-mode of the Virasoro algebra
 and $::$ is the conformal normal ordering.
It is convenient
 for the later use
 to rewrite
 (\ref{eq;nek formula})
 as follows:
\begin{eqnarray}
Z_{\mathcal{N}=1^*}(q,t,\mu)
&=&
\sum_{\lambda}q^{|\lambda|}
\left(
 \prod_{i=1}^{l(\lambda)}
\prod_{n=1}^{\lambda_i}
\displaystyle
 \frac{
 \left[
  n-i+l(\lambda)+\mu
 \right]_{t^{1/2}}
 \left[
  n-i+l(\lambda)-\mu
 \right]_{t^{1/2}}
}{
 \left[
  n-i+l(\lambda)
 \right]_{t^{1/2}}^2
}
\right) \nonumber\\
&&
\times\left(
\prod_{n=1}^{l(\lambda)-1}
\frac{
 \left[
  n-l(\lambda)+\mu
 \right]_{t^{1/2}}^n
 \left[
  n-l(\lambda)-\mu
 \right]_{t^{1/2}}^n
}{
 \left[
  n-l(\lambda)
 \right]_{t^{1/2}}^{2n}
}
\right)\nonumber\\
&&
\times \left[
  \mu
 \right]_{t^{1/2}}^{l(\lambda)}
\times \det_{1\leq i,j \leq l(\lambda)}
\frac{
 \left[
 -\lambda_i+i+\lambda_j-j
 \right]_{t^{1/2}}
}{
 \left[
\mu-\lambda_i+i+\lambda_j-j
 \right]_{t^{1/2}}
},
\label{eq;Z finite}
\end{eqnarray}
where $l(\lambda)$ is the 
 length of $\lambda$.

Let us see that
 (\ref{eq;nek formula})
 is actually written as 
 (\ref{eq;Z and correlation}).
RHS of (\ref{eq;Z and correlation})
 can be computed by using
 two-dimensional 
 free fermions $\psi$ and $\psi^*$,
 which are the fermionization of
 $\varphi$.
The trace in (\ref{eq;Z and correlation})
 becomes the summation over 
 the free fermion states,
 which are 
 related to 
 partitions
 as described 
 in Section \ref{sec;fermion}.
Let $\Psi_{\pm}(x)$ and $\Psi_{\pm}^*(x)$
 be 
 the following 
 dressed free fermions:
\begin{eqnarray}
\Psi_{\pm}(x) &=& \left(
	 \prod_{n=1}^{\mu}\Gamma_{\pm}(t^{-n+\frac{\mu+1}{2}})
		\right)
 \psi(x) \left(
	 \prod_{n=1}^{\mu}\Gamma_{\pm}^{-1}(t^{-n+\frac{\mu+1}{2}})
		\right) \\
&=&
\left(
 \prod_{n=1}^{\mu}(1-t^{\pm(-n+\frac{\mu+1}{2})}x^{\pm 1})^{-1}
\right)
 \psi(x),
\label{eq;Psi}\\
\Psi_{\pm}^*(x) &=& \left(
	 \prod_{n=1}^{\mu}\Gamma_{\pm}(t^{-n+\frac{\mu+1}{2}})
		\right)
 \psi^*(x) \left(
	 \prod_{n=1}^{\mu}\Gamma_{\pm}^{-1}(t^{-n+\frac{\mu+1}{2}})
		\right) \\
&=&\left(
 \prod_{n=1}^{\mu}(1-t^{\pm(-n+\frac{\mu+1}{2})}x^{\pm 1})
\right)
 \psi^*(x),\label{eq;Psi*}
\end{eqnarray}
where
\begin{eqnarray}
\Gamma_{\pm}(z)=\exp(\sum_{k=1}^{\infty}\frac{1}{k}
z^{\pm k} J_{\pm k}) ,
\end{eqnarray}
and we have used the following relations
 in (\ref{eq;Psi}) and (\ref{eq;Psi*}):
\begin{eqnarray}
\Gamma_{\pm}(z)\psi(x)\Gamma_{\pm}^{-1}(z)
 &=& (1-z^{\pm 1} x^{\pm 1})^{-1}\psi(x) \\
\Gamma_{\pm}(z)\psi^*(x)\Gamma_{\pm}^{-1}(z)
&=&
(1-z^{\pm 1} x^{\pm 1})\psi^*(x).
\end{eqnarray}
The functions $\Gamma_{\pm}(z)$
 are introduced in \cite{rpp0}
 to study Schur process, in particular,
 the random plane partition model.
The mode expansions of $\Psi_{\pm}(z)$ and $\Psi_{\pm}^*(z)$
 are
\begin{eqnarray} 
\Psi_{\pm}(z)
 =
 \sum_{r\in \mathbb{Z}+1/2}
 \Psi_{r,\pm} z^{-r-1/2}, \hspace{5mm}
\Psi_{\pm}^*(z)
 =
 \sum_{r\in \mathbb{Z}+1/2}
 \Psi_{r,\pm}^* z^{-r-1/2}.
\end{eqnarray}
By noting (\ref{eq;lambda and nantoka p}),
 we find 
\begin{eqnarray}
&&
\mbox{Tr}\left(
q^{L_0}:\prod_{n=1}^{\mu}
\exp(-i\varphi(t^{-n+\frac{\mu+1}{2}})):
\right) \nonumber\\
&=&
\sum_{\lambda}
q^{|\lambda|}
 \langle \lambda ;0 | 
\left(
\prod_{n=1}^{\mu} \Gamma_{-}^{-1}(t^{-n+\frac{\mu+1}{2}})
\right)\left(
\prod_{n=1}^{\mu} \Gamma_{+}(t^{-n+\frac{\mu+1}{2}})
\right)
| \lambda ;0 \rangle \label{eq;Z and gamma} \\
&=&
\sum_{\lambda}
q^{|\lambda|}
\det_{1\leq i,j \leq l(\lambda)}
  G_{\lambda_i-i \, \lambda_j-j}^{l(\lambda)}
 \label{eq;Gij det},
\end{eqnarray}
 where
\begin{eqnarray}
G_{ij}^l= \langle \phi;-l |
 \Psi^*_{i+1/2, -}\Psi_{-j-1/2, +} 
 | \phi;-l \rangle. 
\end{eqnarray}

The matrix element $G_{ij}^l$
 has the following expression.
\begin{eqnarray}
G_{ij}^l
 &=&
 \mbox{Coeff}_{x^{-i-1}y^{j}}
\langle \phi;-l |
 \Psi^*_{-}(x)\Psi_{+}(y) | \phi;-l \rangle \nonumber\\
 &=&
 \mbox{Coeff}_{x^{-i-1}y^{j}}
\left(
\prod_{n=1}^{\mu}
 \frac{1-t^{n-\frac{\mu+1}{2}}x^{-1}}{1-t^{-n+\frac{\mu+1}{2}}y}
\right)
\left(\frac{x}{y}\right)^l
\frac{1}{x-y} \nonumber \\
&=&
 \mbox{Coeff}_{x^{-i-1-l}y^{j+l}}
\left(
\prod_{n=1}^{\mu}
 \frac{1-t^{n-\frac{\mu+1}{2}}x^{-1}}{1-t^{-n+\frac{\mu+1}{2}}y}
\right)
\frac{1}{x-y}.
 \label{eq;coeff Gij}
\end{eqnarray}
For the case of $G_{\lambda_i-i \, \lambda_j-j}^{l(\lambda)}$,
 which we want to evaluate,
 it is sufficient to expand 
 the function in (\ref{eq;coeff Gij})
 by $x^{-1}$ and $y$
 because
 $\lambda_i-i+l$ is positive for all
 $1\leq i \leq l(\lambda)$.
The function in (\ref{eq;coeff Gij})
 satisfies the following differential equation.
\begin{eqnarray}
&&
(\mu+x\partial_{x}+y\partial_{y})x
\left(
\prod_{i=1}^{\mu}\frac{1-t^{i-\frac{\mu+1}{2}}x^{-1}}
{1-t^{-i+\frac{\mu+1}{2}}y}
\right)
 \frac{1}{x-y} \nonumber\\
&=&   
\left(
\prod_{i=1}^{\mu}\frac{1-t^{i-\frac{\mu+1}{2}}x^{-1}}
{1-t^{-i+\frac{\mu+1}{2}}y}
\right)
\left(
 \sum_{i=1}^{\mu}
 \frac{1}{(1-t^{i-\frac{\mu+1}{2}}x^{-1})(1-t^{-i+\frac{\mu+1}{2}}y)}
\right) \nonumber\\
&=&\sum_{n,m=0}^{\infty}
\sum_{k=1}^{\mu}(-1)^n e_{n,k}h_{m,k}x^{-n}y^m ,
\label{eq;diff Gij}
\end{eqnarray}
 where 
\begin{eqnarray}
e_{n,k}
&=&
e_n(z_1^{-1},\cdots,z_{k-1}^{-1},
0,z_{k+1}^{-1},\cdots,z_{\mu}^{-1}),\\
h_{m,k}
&=&
h_m(z_1,\cdots,z_{\mu},z_{k}).
\end{eqnarray}
$e_n$ is the $n$-th elementary symmetric function,
$h_m$ is the $m$-th complete symmetric function
and $z_i=t^{-i+\frac{(\mu+1)}{2}}$ for $1\leq i \leq \mu$.
From the properties of symmetric functions,
 $e_{n,k}$ and $h_{m,k}$ satisfy
 the following equations.
\begin{eqnarray}
e_{n,k}=\sum_{r=0}^{n} (-t^{k-\frac{\mu+1}{2}})^r
 e_{n-r}(z_1^{-1},\cdots,z_{\mu}^{-1}), 
\hspace{5mm}
h_{m,k}=\sum_{r=0}^{m} (t^{-k+\frac{\mu+1}{2}})^r
 h_{m-r}(z_1,\cdots,z_\mu).
\label{eq;e_k h_k}
\end{eqnarray}
The following equations can be found \cite{macdonald}.
\begin{eqnarray}
e_{n}(z_1^{-1},\cdots,z_{\mu}^{-1}) = \left[
	 \begin{array}{c}
	  \mu \\
	  n
	 \end{array}
	   \right]_{t^{1/2}} &,&
h_{m}(z_1,\cdots,z_{\mu}) =\left[
	 \begin{array}{c}
	  \mu+m-1 \\
	  m
	 \end{array}
	   \right]_{t^{1/2}},
\label{eq;e_n h_n}
\end{eqnarray}
where
\begin{eqnarray}
\left[
 \begin{array}{c}
  \mu \\
  n
 \end{array}
\right]_{t^{1/2}} 
 &=&
 \frac{\left[  \mu \right]_{t^{1/2}}!
 }{
 \left[  n \right]_{t^{1/2}}!
 \left[ \mu- n \right]_{t^{1/2}}!
}, \nonumber\\
\left[ n \right]_{t^{1/2}}!
 &=&
\left[ n \right]_{t^{1/2}}
\left[ n-1 \right]_{t^{1/2}}
\cdots
 \left[ 1 \right]_{t^{1/2}}.
\end{eqnarray}
By using
 (\ref{eq;coeff Gij}),
 (\ref{eq;diff Gij}),
 (\ref{eq;e_k h_k}) and
 (\ref{eq;e_n h_n}),
 we obtain $G_{ij}^l$.
\begin{eqnarray}
G_{ij}^l=
(-1)^{i+l} 
\left[
 \begin{array}{c}
  \mu-n+m-1 \\
  m+l
 \end{array}
\right]_{t^{1/2}}
\left[
 \begin{array}{c}
  \mu +m+l \\
  n+l
 \end{array}
\right]_{t^{1/2}}.
\label{eq;Gij by two coeff}
\end{eqnarray}
By using (\ref{eq;Z finite}),
 (\ref{eq;Gij det}) and
 (\ref{eq;Gij by two coeff}),
 we arrive the equation (\ref{eq;Z and correlation}).

\subsection{Closed expression of the partition function}
$Z_{\mathcal{N}=1^*}(q,t,\mu)$ can be 
 written as a summation of products
 of the skew Schur functions. 
This allows us
 to express $Z_{\mathcal{N}=1^*}(q,t,\mu)$
 in a closed form.

For infinite
 variables $\{x_1,x_2,\cdots\}$
 and $\{y_1,y_2,\cdots\}$,
 there are
 the following equations.
\begin{eqnarray}
\langle \nu | 
\prod_{i=1}^{\infty}\Gamma_{+}(x_i)
 |\lambda \rangle
 &=& s_{\lambda/\nu}(x_i), \\
\langle \lambda |
\prod_{i=1}^{\infty}
 \Gamma_{-}^{-1}(-y_i)
 | \nu \rangle
 &=& s_{\lambda^t/\nu^t}(y_i), 
\end{eqnarray}
 where 
 $s_{\lambda/\nu}$ is the skew Schur function,
 $\lambda$ and $\nu$ are partitions and
 $\lambda^{t}$ is the transpose of $\lambda$.
The following equation
 is an identity \cite{macdonald}.
\begin{eqnarray}
\sum_{\nu,\lambda}
q^{|\lambda|}s_{\lambda/\nu}(x_j)
   s_{\lambda^t/\nu^t}(y_k) 
&=& \prod_{i=1}^{\infty}
\left\{
 (1-q^i)^{-1}
 \prod_{j,k=1}^{\mu}
  (1+q^i x_j y_k)
\right\}.
\end{eqnarray}
Then 
 we find the partition function
 is written as follows:
\begin{eqnarray}
Z_{\mathcal{N}=1^*}(q,t,\mu)
&=&\sum_{\nu,\lambda}
q^{|\lambda|}s_{\lambda/\nu}(t^{-i+\frac{\mu+1}{2}})
   s_{\lambda^{t}/\nu^{t}}(-t^{-i+\frac{\mu+1}{2}})  \nonumber\\
&=& \prod_{i=1}^{\infty}
\left\{
 (1-q^i)^{-1}
 \prod_{j,k=1}^{\mu}
  (1-q^i t^{-j-k+\mu+1})
\right\}\nonumber \\
&=&
\left(
q^{-\frac{1}{24}} \eta(\tau)
\right)^{-\frac{1}{2}(\mu-2)(\mu-1)}
\prod_{n=1}^{\mu-1}\left(
\frac{
\theta_{1\,1}(\frac{n\ln t}{2\pi i}, \tau)
}
{iq^{1/8}(t^{n/2}-t^{-n/2})}
\right)^{\mu-n},
\end{eqnarray}
 where $ \eta(\tau)$
 is the Dedekind eta function
 and $\theta_{1\,1}(\nu,\tau)$
 is the theta function:
\begin{eqnarray}
\eta(\tau)
 &=&
q^{1/24}\prod_{n=1}^{\infty}
(1-q^{n}), \\
\theta_{1\,1}(\nu,\tau)
&=&
-2\exp(\pi i \tau/4) \sin( \pi \tau)
 \prod_{n=1}^{\infty}
 (1-q^n)(1-zq^n)(1-z^{-1}q^n),
\end{eqnarray}
 where
 $q=\exp(2\pi i \tau)$ and $z=\exp(2\pi i \nu)$.

\subsection{Several limits of the gauge theory}
We examine certain limits
 of the gauge theory,
 in particular,
 the limit  to the four-dimensional gauge theory 
 and
 to
 the five-dimensional
 gauge theory without matters.

As $\beta \rightarrow 0$
 ($t\rightarrow 1$),
 $Z_{\mathcal{N}=1^*}(q,t,\mu)$ becomes
 Nekrasov's partition function 
 for the four-dimensional $\mathcal{N}=2^*$ 
 gauge theory which is described in \cite{nek okoun}.
\begin{eqnarray}
Z_{\mathcal{N}=1^*}(q,t,\mu)
&\stackrel{\beta \rightarrow 0}{\longrightarrow}&
\sum_{\lambda} \langle \lambda ;0 | 
q^{L_0}
 \exp(-\sum_{k=1}^{\infty}
 \frac{\mu}{k} J_{-k})
 \exp(\sum_{k=1}^{\infty}
 \frac{\mu}{k} J_{k})
| \lambda ;0 \rangle .
\end{eqnarray}

To obtain the five-dimensional 
 gauge theory without matters,
 we decouple the adjoint matter
 by letting
 $\mu \rightarrow \infty$
 with 
 keeping $q t^{-\mu} = Q$
 fixed.
\begin{eqnarray}
Z_{\mathcal{N}=1^*}(q,t,\mu)&=&
\sum_{\lambda} 
t^{\mu |\lambda|}
\langle \lambda ;0 |
\left(
 \prod_{n=1}^{\mu} \Gamma_{+}(t^{n-1/2})
\right)
 (qt^{-\mu})^{L_0}
\left(
 \prod_{n=1}^{\mu} \Gamma_{-}^{-1}(t^{-n+1/2})
\right)
| \lambda ;0 \rangle \nonumber\\
&\stackrel{\mu\rightarrow \infty ,\, qt^{-\mu}=Q}
{\longrightarrow}&
\langle \phi ;0 |
\left(
 \prod_{n=-\infty}^{-1} \Gamma_{+}(t^{-(n+1/2)})
\right)
 Q^{L_0}
\left(
 \prod_{n=0}^{\infty} \Gamma_{-}^{-1}(t^{-(n+1/2)})
\right)
| \phi ;0 \rangle, 
\label{eq;no matter Nekrasov}
\end{eqnarray}
where we have used
\begin{eqnarray}
t^{\mu |\lambda|}
\stackrel{\mu\rightarrow \infty}{\longrightarrow}
\begin{cases}
1 & \lambda=\phi \\
0 & \lambda\not= \phi.
\end{cases}
\end{eqnarray}
By setting $Q=\beta\Lambda$,
 (\ref{eq;no matter Nekrasov}) becomes Nekrasov's partition function 
 for the five-dimensional gauge theory without matters
 \cite{nek okoun}.

\subsection{$N$ partitions and $\mathcal{N}=1^*$ 
 $SU(N)$ gauge theories}

By embedding $N$ partitions ($N$ component fermions)
 into a single partition (one component fermions),
 we can obtain the partition function for
 the $\mathcal{N}=1^*$ $SU(N)$ gauge theory
 from $Z_{\mathcal{N}=1^*}(q,t,\mu)$.
With the embedding,
 $Z_{\mathcal{N}=1^*}(q,t,\mu)$
 is factorized into two parts:
 One 
 is written as summation over $N$ partitions
and the other is the remaining part. 
We obtain 
 the instanton part of 
 Nekrasov's partition function 
 from the first part
 and
 the perturbative
 prepotential
 from 
 the second part.
For the case of the gauge theory with no matters
 this property is shown in \cite{MNTT1}.

Let $\psi^{(r)}$ and $\psi^{(r)*}$
 be $N$ component fermions labelled by
 $r\in[1,N]$.
We embed the $N$ component fermions
 into one component fermions
 in the standard fashion \cite{Jimbo:1983if};
\begin{eqnarray}
\psi^{(r)}_{s}=\psi_{N(s-\xi_r)}
&&
\psi^{(r)*}_s=\psi_{N(s+\xi_{r})}.
\end{eqnarray}
With this embedding,
 states for the $N$ component fermions
 are mapped to
 states for the one component fermions
 bijectively.
By using the standard correspondence
 between fermion states and partitions,
 for each set of $N$ partitions $\lambda^{(r)}$
 and their charges $p_r$,
 the corresponding 
 single partition $\nu(\lambda^{(r)},p_r)$
 and its charge $P$ are obtained.
The relations between 
 them are such that:
\begin{eqnarray}
\left\{ x_i(\nu(\lambda^{(r)},p_r))+P ; i\geq 1
\right\}
=
\bigcup_{r=1}^{N}
\left\{N(x_{i_r}(\lambda^{(r)})+\tilde{p}_r); i_r \geq 1
\right\} ,
\label{eq;N to 1 map}
\end{eqnarray}
where $x_i(\lambda)=\lambda_i-i+1/2$.
From (\ref{eq;N to 1 map})
the following two equations are found.
\begin{eqnarray}
P&=& \sum_{p_r} p_r, \label{eq;cahrge conv} \\ 
|\nu(\lambda^{(r)},p_r)|
&=& N\sum_{r=1}^{N}|\lambda^{(r)}|
 +\frac{N}{2}\sum_{r=1}^{N}p_r^2
 +\sum_{r=1}^{N}r p_r. \label{eq;1 and N boxes}
\end{eqnarray}
For a set of charges $\{p_r\}$,
 there is a unique partition
 which
 depends only on the charges.
We call it
 a ground partition $\lambda_{GP}$ \cite{MNTT1}:
 $\lambda_{GP}=\nu(\phi^{(r)},p_r)$.

Summation over partitions in
 (\ref{eq;nek formula})
 is expressed as summation over
 $N$ partitions $\lambda^{(r)}$
 and their charges $p_r$.
The charges are restricted to satisfy
 $\sum_{r=1}^{N}p_r=0$
 owing to the charge conservation (\ref{eq;cahrge conv}).
We factorize
 $Z_{\mathcal{N}=1^*}(q,t,\mu)$ 
 into two parts:
 One is written as summation over $N$ partitions
 and the other is the remaining part.
We denote the first part by $Z^{inst}(q,p_r,\mu)$
 and the second part by $Z^{pert}(q,p_r,\mu)$.
\begin{eqnarray}
&& Z_{\mathcal{N}=1^*}(q,t,\mu) \nonumber\\
&=&
\sum_{\{p_r\}}
 \sum_{\{\lambda^{(r)}\}} q^{|\nu(\lambda^{(r)},p_r) |}
 \prod_{(r,i)\not= (s,j)}
\begin{array}[t]{c}
 \displaystyle
\frac{
 \left[N(
  x_i(\lambda^{(r)}) +\tilde{p}_r
  -x_j(\lambda^{(s)})-\tilde{p}_s
) \right]_{t^{1/2}}
}{
 \left[N(
  x_j(\phi^{(r)})
  -x_i(\phi^{(s)})
) \right]_{t^{1/2}}
} \\ 
\times\displaystyle
\frac{
 \left[N(
  \mu/N+ 
  x_j(\phi^{(r)})
  -x_i(\phi^{(s)}) 
 )\right]_{t^{1/2}}
}{
 \left[N(
  \mu/N+ 
  x_i(\lambda^{(r)}) +\tilde{p}_r
  -x_j(\lambda^{(s)})-\tilde{p}_s
) \right]_{t^{1/2}}
} 
\end{array}
 \nonumber\\
&=&
\sum_{\{p_r\}}
\left\{
\begin{array}[c]{cc}
 \sum_{\{\lambda^{(r)}\}}
q^{N\sum_{r=1}^{N}|\lambda^{(r)}|}
 \prod_{(r,i)\not= (s,j)}
&
\displaystyle
\frac{
 \left[N(
  x_i(\lambda^{(r)})+\tilde{p}_r
  -x_j(\lambda^{(s)})-\tilde{p}_s
 )\right]_{t^{1/2}}
}{
 \left[N(
  x_i(\phi^{(r)})+\tilde{p}_r
  -x_j(\phi^{(s)})-\tilde{p}_s
 )\right]_{t^{1/2}}
}
 \\
&
\times\displaystyle
\frac{
 \left[N(
  \mu/N+ 
  x_i(\phi^{(r)})+\tilde{p}_r
  -x_j(\phi^{(s)}) -\tilde{p}_s
 )\right]_{t^{1/2}}
}{
 \left[N(
  \mu/N+ 
  x_i(\lambda^{(r)})+\tilde{p}_r
  -x_j(\lambda^{(s)})-\tilde{p}_s
 )\right]_{t^{1/2}}
}
\end{array}
 \right\} 
\nonumber \\
&&
\hspace{7mm}
\times\left\{
\begin{array}[c]{cc}
 q^{|\lambda_{GP}|}
 \prod_{(r,i)\not= (s,j)}
&
\displaystyle
\frac{
 \left[N(
  x_i(\phi^{(r)}) +\tilde{p}_r
  -x_j(\phi^{(s)})-\tilde{p}_s
) \right]_{t^{1/2}}
}{
 \left[N(
  x_i(\phi^{(r)})
  -x_j(\phi^{(s)})
 )\right]_{t^{1/2}}
}
 \\ 
&
\times\displaystyle
\frac{
 \left[N(
  \mu/N+ 
  x_i(\phi^{(r)})
  -x_j(\phi^{(s)}) 
) \right]_{t^{1/2}}
}{
 \left[N(
  \mu/N+ 
  x_i(\phi^{(r)}) +\tilde{p}_r
  -x_j(\phi^{(s)})-\tilde{p}_s
) \right]_{t^{1/2}}
}
\end{array}
\right\}
 \nonumber\\
&=&
\sum_{\{p_r\}}Z^{inst}(q,p_r,\mu)\cdot Z^{pert}(q,p_r,\mu),
 \label{eq;inst and pert}
\end{eqnarray}
where
 we have used (\ref{eq;1 and N boxes}) in the third line.

With the following identification,
 we can interpret
 $Z^{inst}(q,p_r,\mu)$ and $Z^{pert}(q,p_r,\mu)$
 respectively
 as
 the instanton part and the perturbative part
 of
 a five-dimensional generalization 
 of Nekrasov's partition function
 for $SU(N)$ gauge theory.
\begin{eqnarray}
\tilde{p}_r = a_r/\hbar ,\hspace{3mm}
 t= \exp(-\frac{\beta \hbar}{N}) ,\hspace{3mm}
  \mu=\frac{N m}{\hbar} , \hspace{3mm}
\frac{\ln q}{N} = 2\pi i  \tau , \hspace{3mm}
\tau= \frac{4\pi i}{g_{YM}^2}.
 \label{eq;su n identifications}
\end{eqnarray}
\begin{eqnarray}
Z^{inst}(q,p_r,\mu)=
\sum_{\lambda^{(r)}} e^{2 \pi i \tau\sum_{r=1}^{N}|\lambda^{(r)}|}
\prod_{(r,i)\not= (s,j)}
\begin{array}[t]{l}
\displaystyle
\frac{
 \sinh(\frac{\beta\hbar}{2}( 
 a_{rs}/\hbar +\lambda_i^{(r)} -\lambda_j^{(s)}+j-i))
}{ 
 \sinh(\frac{\beta\hbar}{2} (
 a_{rs}/\hbar+j-i))
 }
\\ 
\times
\displaystyle
\frac{
 \sinh(\frac{\beta\hbar}{2}( 
 (m+a_{rs})/\hbar+j-i))
 }{
 \sinh(\frac{\beta\hbar}{2}( 
 (m+ a_{rs})/\hbar +\lambda_i^{(r)} -\lambda_j^{(s)}+j-i))
}.
\end{array}
\end{eqnarray}

\begin{figure}[hbt]
\begin{center}
\psfrag{12}{\hspace{-2mm}$(3,2)$}
\psfrag{23}{\hspace{-2mm}$(2,1)$}
\psfrag{13}{\hspace{-2mm}$(3,1)$}
 \includegraphics{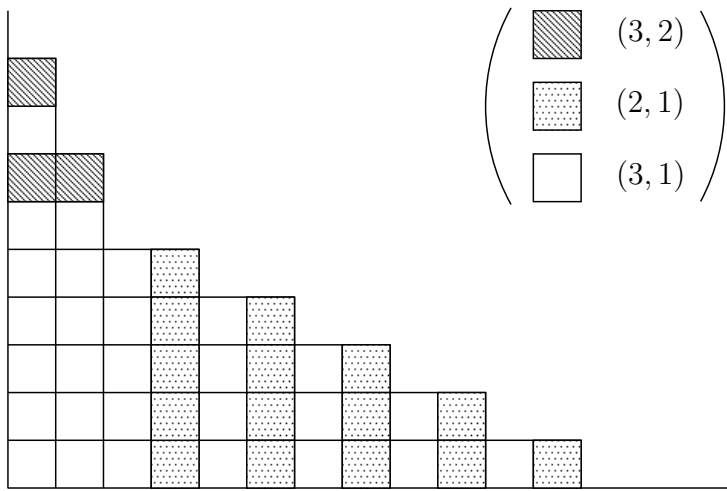}
 \caption{\textit{The example of bicolouring 
 in the case of $N=3$, $T_1=2$ and $T_2=5$.
In this case
 the ground partition is
 $\lambda_{GP}=(9,7,5,5,4,4,3,3,2,2,1,1)$.
The upper-right parenthesis in the figure
 means the map between the shading patterns of boxes
 and the $(r,s)$-colours of boxes.
}}
 \label{fig;bicolour N=3}
\end{center}
\end{figure}

To see the emergence of the perturbative 
 prepotential from $Z^{pert}(q,p_r,\mu)$,
 we rewrite it as follows:
\begin{eqnarray}
Z^{pert}(q,p_r,\mu)
&=&
q^{|\lambda_{GP}|}
\prod_{i\not= j}
\frac{
\left[
 \lambda_{GP\, i} -\lambda_{GP\, j} 
 +j-i
\right]_{t^{1/2}}
}{
\left[
 j-i
\right]_{t^{1/2}}
}
\frac{
 \left[
 \mu +j-i
 \right]_{t^{1/2}}
}{
\left[
 \mu +\lambda_{GP\, i} -\lambda_{GP\, j} 
 +j-i
\right]_{t^{1/2}}
}  \nonumber\\
&=&
q^{|\lambda_{GP}|}
\prod_{(i,j)\in \lambda_{GP}}
\frac{
 \left[
 h(i,j)+\mu
 \right]_{t^{1/2}}
 \left[
 h(i,j)-\mu
 \right]_{t^{1/2}}
}{ 
\left[
 h(i,j)
 \right]_{t^{1/2}}^2
} \label{eq;pert hock}
\end{eqnarray}
 where $h(i,j)$ is the hook length for the box $(i,j)\in \lambda_{GP}$:
 $h(i,j)=\lambda_{GP\,i}-i+\lambda_{GP\,j}^t-j$.
The hook length are
 easily obtained
 by bicolouring each box
 in $\lambda_{GP}$ \cite{MNTT1}.
We bicolour each box $(i,j)\in\lambda_{GP}$
 by $(r,s)$-colour ($r,s\in [1,N]$) when
 $i$ and $j$ satisfy
 $x_i(\lambda_{GP})\in N\mathbb{Z}+r-\frac{1}{2}$ and 
 $x_j(\lambda_{GP}^t)\in N\mathbb{Z}-s+\frac{1}{2}$.
An example of the bicolouring 
 for the $SU(3)$ ground partition is shown in
 Figure \ref{fig;bicolour N=3}.
The collection of $(r,s)$-coloured box
 forms a partition,
 which we
 call $(r,s)$-coloured partition $\lambda_{rs}$.
The partition is
$\lambda_{rs}=(T_{rs},T_{rs}-1,\cdots,1)$,
 where
\begin{eqnarray}
T_{rs}= \sum_{n=N-r+1}^{N-s} T_n,
 \label{eq;Trs}
\end{eqnarray}
 and $T_n$ are defined as differences of the charges:
 $T_n=p_{N-n+1}-p_{N-n}$.
The box $(i,j)\in \lambda_{rs}$
 corresponds to a box in $\lambda_{GP}$
 and its hook length is $N(T_{rs}-i-j+2)+r-s$.
Hence there are $T_{rs}-n$ boxes
 whose
 hook length are $Nn+r-s$. 
As $\hbar\rightarrow 0$,
 each part of $Z^{pert}(q,p_r,\mu)$
 become
\begin{eqnarray}
|\lambda_{GP}| 
&\stackrel{\hbar \rightarrow 0}{\longrightarrow}&
\frac{1}{\hbar^2}
\sum_{r>s}\frac{a_{rs}}{2}  +\mathcal{O}(\hbar^{-1}),
\nonumber \\
\sum_{(i,j)\in \lambda_{GP}}
 \ln[h(i,j)+\mu]_{t^{1/2}} 
&\stackrel{\hbar \rightarrow 0,\,\beta \rightarrow \infty}
{\longrightarrow}&
\frac{\beta}{\hbar^2}
\sum_{r>s}
\left(
 \frac{1}{12}(a_{rs}+m)^3
 -\frac{1}{12}(3a_{rs}m^2+m^3)
\right) +\mathcal{O}(\hbar^{-1}), \nonumber \\
\sum_{(i,j)\in \lambda_{GP}}
 \ln[h(i,j)-\mu]_{t^{1/2}}
&\stackrel{\hbar \rightarrow 0,\,\beta \rightarrow \infty}
{\longrightarrow}&
\frac{\beta}{\hbar^2}
\sum_{r>s}
\left(
 \frac{1}{12}(a_{rs}-m)^3
+\frac{1}{12}(3a_{rs}m^2-m^3)
\right) 
+\frac{i}{\hbar^{2}} \mathcal{O}(T_{rs})
 +\mathcal{O}(\hbar^{-1}),
 \nonumber \\
\sum_{(i,j)\in \lambda_{GP}}
 \ln[h(i,j)]_{t^{1/2}} 
&\stackrel{\hbar \rightarrow 0,\,\beta \rightarrow \infty}
{\longrightarrow}&
\frac{\beta}{\hbar^2}
\sum_{r>s}
\left(
 \frac{1}{12}a_{rs}^3
\right) +\mathcal{O}(\hbar^{-1}),
\end{eqnarray}
where we assumed $a_{r+1}-a_r > \mu$
 and $\beta\gg1$ 
to compare the dimension counting argument
 in Section \ref{sec;dim count adj}.
We can obtain
 the perturbative prepotential for 
 the five-dimensional $\mathcal{N}=1^*$
 $SU(N)$ gauge theory
 from $Z^{pert}(q,p_r,\mu)$.
\begin{eqnarray}
\ln Z^{pert}(q,p_r,\mu)
&\stackrel{\hbar \rightarrow 0 ,\,
\beta \rightarrow \infty}{\longrightarrow}&
\frac{-1}{\hbar^2}
\mathcal{F}^{pert}_{adj}(a_r;\beta ,- 2N\pi i \tau ,0, m) 
-\frac{\beta N(N-1)}{12\hbar^2}m^3
+\frac{i}{\hbar^{2}} \mathcal{O}(T_{rs})
+\mathcal{O}(\hbar^{-1}), \nonumber\\
\label{eq;z pert and f pert}
\end{eqnarray}
where $\mathcal{F}^{pert}_{adj}(a_r;\beta ,- 2N \pi i \tau ,0, m)$
 is given in (\ref{eq;pre pert adj}).

\section{Statistical models of partitions and Polyhedrons}
\label{sec;rpp}

Nekrasov's partition function
 for the five-dimensional $U(1)$ gauge theory
 with no matters
 is expressed as
 a partition function
 of a random plane partition model \cite{MNTT1}.
For the case of
 the partition function
 $Z_{\mathcal{N}=1^*}(q,t,\mu)$,
 we can rewrite it
 as 
 a partition function
 of a statistical model of partitions.
Each ground partition
 corresponds to
 a ground state of the model.
We can reproduce
 the polyhedron
 $\widetilde{\mathcal{P}}_{adj}^c$
 from the ground state.
The reproduction 
 of the polyhedron in the case of 
 the gauge theory with no matter
 is described in \cite{MNNT}.

For partitions $\lambda$ and $\rho$,
 the following equations hold.
\begin{eqnarray}
\langle \lambda | \Gamma_{+} (1)
 | \rho \rangle 
 &=&
 \begin{cases}
  1
   & \mbox{if $\lambda \prec \rho$} \\
  0 
   & \mbox{if $\lambda \not\prec \rho$}
  \end{cases} \label{eq;gamma_p p}\\
\langle \lambda | \Gamma_{-}^{-1} (1)
 | \rho \rangle 
 &=&
 \begin{cases}
  (-1)^{|\lambda | - |\rho|} 
   & \mbox{if $\lambda^{t} \succ \rho^{t}$} \\
  0 
   & \mbox{if $\lambda^{t} \not\succ \rho^{t}$}
  \end{cases}
\label{eq;gamma_m p}
\end{eqnarray}
where $\lambda \succ \rho$
 means
 $\lambda_1 \geq \rho_1 \geq
  \lambda_2\geq \rho_2 \geq \cdots$.
By using (\ref{eq;gamma_p p}) and (\ref{eq;gamma_m p}),
 $Z_{\mathcal{N}=1^*}(q,t,\mu)$ is written as
 a partition function of 
 a statistical model of partitions.
\begin{eqnarray}
Z_{\mathcal{N}=1^*}(q,t,\mu)
&=&
\sum_{\lambda} \langle \lambda ;0 | 
t^{-\mu L_0}
 \left(
  t^{L_0}\Gamma_{+}(1)
 \cdots
  t^{L_0}\Gamma_{+}(1) 
 \right)
(qt^{-\mu})^{L_0}
 \left(
  t^{L_0}\Gamma_{-}^{-1}(1)
 \cdots
  t^{L_0}\Gamma_{-}^{-1}(1)
 \right)
| \lambda ;0 \rangle \nonumber \\
&=&
\sum_{\pi}
\left(
\prod_{n=-\mu+1}^{\mu-1}
 t^{|\pi(n)|}
\right)
 (-qt^{-\mu})^{|\pi(0)|}
 (-t^{-\mu+1})^{|\pi(\mu)|}
\end{eqnarray}
where 
 $\pi$ is a sequence of partitions $\pi(n)$
$(-\mu \leq n \leq \mu)$, which
 satisfy
 the following relations,
\begin{eqnarray}
 \pi(-\mu)\prec \pi(-\mu+1) \prec \cdots 
 \prec \pi(0), \hspace{3mm}
 \pi(0)^{t} \succ \pi(1)^{t} \succ \cdots
  \succ \pi(\mu)^{t}, \hspace{3mm}
 \pi(-\mu)=\pi(\mu). \label{eq;interlace t}
\end{eqnarray}

$\pi$ can be seen as 
 a transposed version of plane partitions.
In the case of a plane partition $\nu$,
 it is written as a sequence of partitions
 $\nu(n)$ satisfying
 $\cdots \prec \nu(-2) \prec \nu(-1) \prec \nu(0)
 \succ \nu(1) \succ \nu(2) \succ \cdots$
 and
 is seen 
 as evolutions of $\nu(n)$
 along $n$ from $n=-\infty$ to $n=+\infty$ \cite{rpp0}.
In the case of $\pi$,
 it can be seen as
 a periodic evolution of partition $\pi(n)$ 
 along $n$.
However, by the effect of 
 $(q t^{-\mu})^{|\pi(0)|}$
 and $t^{\mu|\pi(\mu)|}$,
 the evolution becomes more complex than the one for 
 plane partitions.
The effect of $(q t^{-\mu})^{|\pi(0)|}$
 is moving $\pi(0)$ ahead by 
 $-\ln q/(g_{st})- \mu$ steps
 and the effect of $t^{\mu|\pi(\mu)|}$
 is moving $\pi(\mu)$ behind by
 $\mu$ steps.
Then the period of the evolution
 becomes $-\ln q /(\beta\hbar)$.
This is equal to the period of 
 $\widetilde{\mathcal{P}}_{adj}$. 

For the ground partition $\lambda_{GP}$,
 we can define 
 the following set $P_{GP}$
 of sequences of partitions 
 satisfying (\ref{eq;interlace t}).
\begin{eqnarray}
P_{GP}
=
 \left\{
 \pi | \pi(0)= \lambda_{GP}
 \right\}.
\end{eqnarray}
There exists only one sequence of partitions,
 which denoted by $\pi_{GPP}$,
 whose number of boxes 
 achieves the minimum
 in $P_{GP}$.
\begin{eqnarray}
^\exists\pi_{GPP} \in P_{GP} ,\hspace{3mm}
s.t. \hspace{3mm}
|\pi_{GPP}| \leq |\pi|,\hspace{3mm}
 ^\forall\pi \in P_{GP}
\end{eqnarray}
$\pi_{GPP}$ is the ground state of the 
 statistical model.
The explicit form of $\pi_{GPP}$
 is as follows:
\begin{eqnarray}
\pi_{GPP\,i}(n)=
\begin{cases}
 \max\{
\lambda_{GP\, i}-n,\, \lambda^{\mu}_i\}
& \mbox{for }(n\geq 0) \\
 \max\{
\lambda_{GP\, i+n},\, \lambda^{\mu}_i\}
& \mbox{for }(n< 0),
\end{cases}
\end{eqnarray}
where  $\lambda^{\mu}$ is the
 following partition:
\begin{eqnarray}
\lambda^{\mu}_i
 =
\max\{\lambda_{GP\, i+\mu}, \lambda_{GP\, i}-\mu
\}.
\end{eqnarray}

To make a relation
 between $\pi_{GPP}$
 and $\widetilde{\mathcal{P}}_{adj}^c$,
 we define the following sequence of partitions
 $\Upsilon(n)$,
 $(\frac{\ln q}{2\beta\hbar}-\frac{\mu}{2}
< n \leq 
-\frac{\ln q}{2\beta\hbar}-\frac{\mu}{2})$:
\begin{eqnarray}
\Upsilon(n)_i
= 
\begin{cases}
\pi_{GPP}(0)_i & \mbox{if }
 \frac{\ln q}{2\beta\hbar}-\frac{\mu}{2}
 < n \leq -\mu \\
\max\left\{
\pi_{GPP}(n)_i, \, \pi_{GPP}(\mu+n)_i
\right\}  & \mbox{if }-\mu < n \leq 0 \\
\pi_{GPP}(0)_i & \mbox{if }
 0 < n \leq -\frac{\ln q}{2\beta\hbar}-\frac{\mu}{2}.
\end{cases}
\end{eqnarray}
Let $M^{\vee}_N$
 be a lattice
 generated by
 $e_1^*$,
 $\frac{1}{N}e_2^*$
 and $e_3^*$.
We can reproduce 
 all points 
 $m+(K_{0\,N}\cap K_{1\,0}\cap K_{1\,N})
 \in\widetilde{\mathcal{P}}_{adj}^c\cap M^{\vee}_N$
 from
 the partitions
 $\Upsilon(n)$
 ($\frac{\ln q}{2\beta\hbar}-\frac{\mu}{2}
 <n\leq -\frac{\ln q}{2\beta\hbar}-\frac{\mu}{2}$)
 bijectively
 by the following map:
\begin{eqnarray}
m=
\begin{cases}
n e_1^*+\frac{1}{N}(-\mu+j-i+1)e_2^*
 +(\mu-n+i-1)e_3^*
&
\mbox{for }
 \frac{\ln q}{2\beta\hbar}-\mu/2
 < n \leq -\mu \\
n e_1^*+\frac{1}{N}(n+j-i+1)e_2^*
 +(-n+i-1)e_3^*
&
\mbox{for }
 -\mu < n \leq 0  \\
n e_1^*+\frac{1}{N}(j-i+1)e_2^*
 +(i-1)e_3^*
&
\mbox{for }
0 < n \leq -\frac{\ln q}{2\beta\hbar}-\mu/2,
\end{cases}\nonumber
\end{eqnarray}
 where
 $(i,j)\in\Upsilon(n)$.

\appendix
\section{Partitions and fermions}
 \label{sec;fermion}

A partitions $\lambda$ is a sequence of non-negative integers:
 $\lambda=(\lambda_{1}, \lambda_{2}, \cdots )$.
$l(\lambda)$ is
 the number of non-zero integers
 in $\lambda$
 and
 called the length of $\lambda$.

Let $\psi(z) 
 =
 \sum_{r\in \mathbb{Z}+1/2} \psi_r z^{-r-1/2}$
 and 
 $\psi^*(z)
 =
 \sum_{r\in \mathbb{Z}+1/2} \psi^*_r z^{-r-1/2}$
 be two-dimensional free fermions
 with anti-commutation relations
 $\{
 \psi_r,\psi_s^*
 \}= \delta_{r+s,0}$.
The vacuum state $|\phi\rangle$ and its dual state
 $\langle \phi |$
 are defined
 as follows:
\begin{eqnarray}
\psi_{r} | \phi\rangle=0, \hspace{3mm}
\psi_{r}^* | \phi\rangle=0 
\hspace{5mm}
\mbox{for $r>0$.} \nonumber \\
\langle \phi |\psi_{r} =0, \hspace{3mm}
\langle \phi |\psi_{r}^* =0 
\hspace{5mm}
\mbox{for $r<0$.}
\end{eqnarray}
$J_k$ are modes of the $U(1)$ current 
 for the free fermions 
 and 
 written as follows:
\begin{eqnarray} 
J_{k}= \sum_{r\in \mathbb{Z}+1/2}
 :\psi_{k-r}\psi^*_{r}: \, .
\end{eqnarray}

There is a bijective map
 between 
 states of the fermions
 and partitions.
\begin{eqnarray}
| \lambda ;p\rangle &=& (\psi_{-x_1(\lambda)-p} 
 \cdots \psi_{-x_l(\lambda)-p})
 (\psi^*_{x_l(\phi)+p} \cdots \psi^*_{x_1(\phi)+p})
 |\phi; p \rangle \\
\langle \lambda ;p | &=& \langle \phi ;p |
 (\psi_{-x_1(\phi)-p} \cdots \psi_{-x_l(\phi)-p})
(\psi^*_{x_l(\lambda)+p} 
 \cdots \psi_{x_1(\lambda)+p}),
\end{eqnarray}
where $\lambda$ is a partition,
 $p$ is its charge,
 $l=l(\lambda)$ and
\begin{eqnarray}
| \phi ;p \rangle = \begin{cases}
		     \psi_{-p+1/2}\cdots \psi_{-1/2} |\phi;0 \rangle
		     & \mbox{if $p\geq 0$} \\
		     \psi^*_{p+1/2}\cdots \psi^*_{-1/2} |\phi;0 \rangle
		     & \mbox{if $p < 0$} 
		     \end{cases}
\end{eqnarray}

$| \lambda ; 0 \rangle$ and $\langle \lambda ;0 |$
 are written as follows:
\begin{eqnarray}
| \lambda ; 0 \rangle &=& (\psi_{-x_1(\lambda)} 
 \cdots \psi_{-x_l(\lambda)})
 |\phi ;-l(\lambda) \rangle  \\
\langle \lambda ;0 |
&=& \langle \phi ;-l(\lambda) |
(\psi^*_{x_l(\lambda)} 
 \cdots \psi_{x_1(\lambda)}).
 \label{eq;lambda and nantoka p}
\end{eqnarray}

\subsection*{Acknowledgements}
We thank to 
 T.~Nakatsu,
 T.~Tamakoshi and T.~Maeda
 for a useful discussion.
This work is supported in part
 by JSPS Research Fellowships
 for Young Scientists.


\end{document}